\documentclass[10pt,journal,compsoc]{IEEEtran}
\usepackage{amsmath,amsfonts}
\usepackage{algorithmic}
\usepackage[linesnumbered, ruled]{algorithm2e}
\usepackage{array}
\usepackage[caption=false,font=footnotesize,labelfont=rm,textfont=rm]{subfig}
\usepackage{textcomp}
\usepackage{stfloats}
\usepackage{url}
\usepackage{verbatim}
\usepackage{graphicx}
\usepackage{cite}

\usepackage{amssymb}
\usepackage{color}
\usepackage{multirow}
\usepackage{tabularx}
\usepackage{amssymb}
\usepackage{pifont}
\begin{document}

\title{{\color{black}Low-altitude UAV Friendly-Jamming for Satellite-Maritime Communications via Generative AI-enabled Deep Reinforcement Learning}\\
}

\author{Jiawei Huang,
        Aimin Wang,
        Geng Sun,~\IEEEmembership{Senior Member,~IEEE,}
    Jiahui~Li, 
    Jiacheng Wang, \\
    Dusit Niyato,~\IEEEmembership{Fellow,~IEEE,}
    Victor C. M. Leung,~\IEEEmembership{Life Fellow,~IEEE}
\thanks{
This study is supported in part by the National Natural Science Foundation of China (62272194, 62471200), in part by the Science and Technology Development Plan Project of Jilin Province (20250101027JJ), in part by Seatrium New Energy Laboratory, Singapore Ministry of Education (MOE) Tier 1 (RT5/23 and RG24/24), the Nanyang Technological University (NTU) Centre for Computational Technologies in Finance (NTU-CCTF), and the Research Innovation and Enterprise (RIE) 2025 Industry Alignment Fund - Industry Collaboration Projects (IAF-ICP) (Award I2301E0026), administered by Agency for Science, Technology and Research (A*STAR), in part by the Postdoctoral Fellowship Program of China Postdoctoral Science Foundation (GZC20240592), in part by China Postdoctoral Science Foundation General Fund (2024M761123), and in part by Graduate Innovation Fund of Jilin University (2025CX213) (Corresponding authors: Geng Sun and Jiahui Li.)\\
\indent Jiawei Huang, Aimin Wang and Jiahui Li are with the College of Computer Science and Technology, Jilin University, Changchun 130012, China, and Key Laboratory of Symbolic Computation and Knowledge Engineering of Ministry of Education, Jilin University, Changchun 130012, China (E-mails: huangjiawei97@foxmail.com, wangam@jlu.edu.cn, lijiahui@jlu.edu.cn).\\
\indent Geng Sun is with the College of Computer Science and Technology, Key Laboratory of Symbolic Computation and Knowledge Engineering of Ministry of Education, Jilin University, Changchun 130012, China, and also with the College of Computing and Data Science, Nanyang Technological University, Singapore 639798 (E-mail: sungeng@jlu.edu.cn).\\
\indent Jiacheng Wang and Dusit Niyato are with the College of Computing and Data Science, Nanyang Technological University, Singapore 639798 (E-mails: jiacheng.wang@ntu.edu.sg, dniyato@ntu.edu.sg).\\
\indent Victor C. M. Leung is with the Artificial Intelligence Research Institute, Shenzhen MSU-BIT University, Shenzhen 518115, China, with the College of Computer Science and Software Engineering, Shenzhen University, Shenzhen 518060, China, and also with the Department of Electrical and Computer Engineering, The University of British Columbia, Vancouver V6T 1Z4, Canada (e-mail: vleung@ieee.org).}
}



\IEEEtitleabstractindextext{
\begin{abstract}
\textcolor{black}{Low Earth orbit (LEO) satellites can be used to assist maritime wireless communications for wide-area data transmission. However, the extensive coverage of LEO satellites, combined with the openness of channels, can cause the communication process to suffer from security risks. This paper presents a LEO satellite-maritime communication system assisted by low-altitude unmanned aerial vehicle (UAV) friendly-jamming to ensure data security at the physical layer. Since such a system requires balancing the conflicting performance metrics of secrecy rate and energy consumption of the UAV to meet evolving scenario demands, we formulate a secure satellite-maritime communication multi-objective optimization problem (SSMCMOP). In order to solve the dynamic and long-term optimization problem, we reformulate it into a Markov decision process. We then propose a transformer-enhanced soft actor-critic (TransSAC) algorithm, which is a generative artificial intelligence-enabled deep reinforcement learning approach to solve the reformulated problem, thus capturing strong temporal correlations and diversely exploring weights. Simulation results demonstrate that the TransSAC algorithm outperforms comparative approaches and algorithms, maximizing the secrecy rate while effectively minimizing the energy consumption of the UAV. Moreover, the results identify more suitable constraints for the system.}
\end{abstract}

}

%
\maketitle

\IEEEdisplaynontitleabstractindextext

\IEEEpeerreviewmaketitle

\IEEEraisesectionheading{\section{Introduction}
\label{sec:introduction}}

\IEEEPARstart{I}{n} recent years, with the rapid expansion of the maritime economy, the importance of maritime communications has increased significantly~\cite{Yuan2024}. {\color{black}A reliable and stable communication network is essential to ensure the efficiency of maritime operations~\cite{Zeng2025}.} However, the challenges of deploying foundational communication infrastructures, combined with the complexity and variability of the maritime channels, may result in lower transmission rates for maritime networks than terrestrial cellular networks~\cite{Zhang2024}. In this case, satellites, with extensive coverage capabilities, are increasingly being utilized for maritime data transmission, thereby facilitating the effective exchange of data from vessels at sea~\cite{Zhou2023}. In particular, the low Earth orbit (LEO) satellites, operating closer to the Earth, have higher stability for enhancing communication performance~\cite {Pokhrel2024}. However, the open channels and extensive coverage of LEO satellites make them vulnerable to unauthorized access and potential eavesdropping by malicious users, which may pose security risks~\cite{Yue2023}. Although the traditional cryptography-based methods can cope with the security risks in some cases, they require frequent data encoding and decoding, and complex key distribution and management, which pose challenges in the energy-limited maritime environment when large amounts of data are transmitted~\cite{Wu2018}.

\par {\color{black}In this case, physical layer security (PLS) can dynamically adjust security mechanisms, making it a suitable way to ensure secure communication in the LEO satellite-maritime networks~\cite{Wang2019}.} For example, through intelligent beamforming methods, LEO satellites can focus signals on target vessels by optimizing the directionality and power distribution of the transmitting antenna, thereby reducing the probability of illegitimate users acquiring signals~\cite{Zheng2022}. However, for high-speed moving LEO satellites, real-time computation and adjustment of beamforming parameters place high demands on the hardware, leading to a greater computational burden~\cite{Xv2024}. As such, low-altitude platforms are used to introduce friendly-jamming signals to enhance PLS.

\par {\color{black}Common low-altitude platforms include unmanned aerial vehicles (UAVs) and electric vertical take-off and landing (eVTOL) aircraft~\cite{Huang2024b}. In complex maritime environments, UAVs, with low costs, adjustable flight altitudes, and flexible mobility, are particularly suitable for overcoming movement and operation limitations imposed by sea terrain and obstacles~\cite{Qian2022}.} For example, the authors in~\cite{Huang2025} utilized UAVs to form a virtual antenna array as a jammer to send jamming signals to an illegitimate vessel, thus protecting legitimate data signals against being eavesdropped. The authors in~\cite{DangNgoc2022} introduced a cooperative jamming scheme for UAV-assisted networks to enhance security by regulating the position of UAVs. However, the aforementioned works overlook the mobility of vessels, which is a crucial factor for maintaining reliable connectivity in practical scenarios~\cite{Wang2024}. {\color{black}In such dynamic conditions, the original solutions become ineffective, as vessel movement degrades previously stable links. Moreover, continuous real-time computation during the operation creates computational overhead that affects the responsiveness of the system. Therefore, to achieve the LEO satellite-maritime PLS communication system via UAVs, we need to focus on the movement of vessels and the temporal dynamics of maritime channels.}

\par {\color{black}Implementing such systems encounters several challenges. \textit{First}, the movements of vessels are dynamic as they traverse the ocean in varying directions, constantly changing their relative positions in communication networks. Similarly, LEO satellites experience rapid motion, continuously altering their positions relative to the vessels. In this case, traditional offline methods (e.g., convex optimization and evolutionary computation) become ineffective in such a dynamic condition~\cite{Yuan2020},~\cite{Wang2023}, while deep reinforcement learning (DRL) algorithms integrate the feature extraction capabilities of deep learning and the decision-making capabilities of reinforcement learning, demonstrating remarkable adaptability in dynamic environments~\cite{Liao2024},~\cite{Shao2024}. \textit{Second}, when controlling UAVs for jamming, the potential impact on legitimate users needs to be considered to ensure their normal communications. As such, the transmit power of the UAV should be carefully optimized based on user requirements. Meanwhile, UAVs need to frequently adjust their positions to optimize system performance, which increases their energy consumption. Thus, secure communications and energy consumption are conflicting and require balancing. Additionally, since the relative importance of these objectives can change depending on the specific scenarios, existing methods (\textit{i.e.},~\cite{Cai2020}) that prioritize one goal and constrain others become inapplicable for this case. \textit{Finally}, our considered scenario is time-sensitive and involves large-scale decision making, which requires solutions that can adapt quickly while remaining computationally efficient, further increasing the complexity. Thus, an innovative approach, different from previous works, is needed to jointly optimize multiple conflicting objectives and efficiently respond to the dynamic maritime conditions.}

\par Accordingly, we formulate a multi-objective optimization problem (MOP), and propose a generative AI (GenAI)-enabled DRL approach to solve the problem. The main contributions of this work are summarized as follows.

\begin{itemize}
{\color{black}
\item \textbf{\textit{Low-altitude UAV Friendly-Jamming for the LEO Satellite-Maritime Communication System:}} We present an LEO satellite-maritime communication system assisted by low-altitude UAV friendly-jamming, where an LEO satellite sends data signals to a legitimate vessel within range, and a UAV sends jamming signals to a potential eavesdropping vessel. This system is the first to holistically consider the movements of LEO satellites and vessels while designing an adaptive UAV friendly-jamming mechanism that ensures real-time maritime communication security.

\item \textbf{\textit{Multi-objective Optimization Problem Formulation:}} The security performance and energy consumption of the system are in conflict due to shared decision variables. To balance these key performance metrics, we formulate a secure satellite-maritime communication multi-objective optimization problem (SSMCMOP) to jointly maximize the secrecy rate and minimize the energy consumption of the UAV. However, the problem is an NP-hard and long-term optimization problem, making it more complex to be solved.

\item \textbf{\textit{GenAI-enabled DRL Approach:}} Conventional DRL algorithms confront the challenges of large solution spaces and strong temporal correlation when addressing the formulated SSMCMOP. In this case, we first reformulate the problem into a Markov decision process (MDP). We then propose a transformer-enhanced soft actor-critic (TransSAC) algorithm, which is a GenAI-enabled DRL approach to solve the problem. Specifically, the TransSAC algorithm captures temporal dependencies through encoded sequences and accelerates learning policies via parallel processing, while diversely exploring weights to effectively balance multiple optimization objectives.

\item \textbf{\textit{Simulations and Performance:}} Simulation results indicate that the proposed UAV-assisted approach can achieve secure and energy-efficient LEO satellite-maritime communications, significantly outperforming the non-UAV approach. Moreover, comparative results illustrate that TransSAC outperforms other conventional DRL algorithms, achieving the maximum secrecy rate with the minimum energy consumption of the UAV. In addition, we identify optimal constraint values for the formulated MDP, thereby enhancing the performance of algorithms.
}
\end{itemize} 

\par The rest of this paper is organized as follows: Section \ref{sec:Related work} reviews the related work. Section \ref{sec:models_and_formulataion} introduces the models and preliminaries. Section \ref{sec:problem formulation and analysis} formulates the SSMCMOP. The GenAI-enabled DRL approach is proposed in Section \ref{sec:algorithm}. Section \ref{sec:simulation-results-and-analysis} illustrates the simulation results. {\color{black}Section \ref{sec:Discussion} presents some discussions and Section \ref{sec:conclusion} summarizes the overall work.}

%
\section{Related work} 
\label{sec:Related work}

\par In this section, we review the related works associated with LEO satellite maritime communications, security strategies, and optimization approaches. 

\subsection{LEO Satellite-Maritime Communications}

\par {\color{black}The rapid advancement of maritime wireless communications has attracted much attention, while deployment problems in practice pose a considerable challenge to transmission rates~\cite{He2025}.} The satellites, with wide coverage capabilities, serve as valuable auxiliary equipment to facilitate the effective exchange of centralized data and information between vessels at sea~\cite{Alqurashi2023}. {\color{black}For example, the authors in~\cite{Dai2025} proposed an energy-efficient multi-access edge computing scheme for heterogeneous satellite-maritime networks to enhance the perception and offloading endurance of UAVs.} Moreover, the authors in~\cite{Wu2024} proposed an intelligent spectrum-sharing scheme for the satellite-maritime integrated network to optimize throughput and spectral efficiency. However, the significant latency caused by long-distance satellite transmission can negatively impact the efficiency of communications. 

\par LEO satellites, with their proximity to Earth, significantly reduce latency and improve data transmission efficiency. Moreover, LEO satellites can be deployed in satellite constellations to achieve seamless global coverage, allowing them to be gradually applied in maritime communications. For example, the authors in~\cite{Hu2024} proposed an LEO satellite-assisted shore-to-vessel network to achieve end-to-end communications. {\color{black}The authors in~\cite{Senadhira2025} considered a UAV-assisted LEO satellite communication network to provide coverage for low-end maritime users.} Note that the open channels and extensive coverage of satellites make them susceptible to eavesdropping during data transmission. However, the aforementioned works focus mainly on communication efficiency, which ignores the potential security risks.

\subsection{Security Strategies}

\par As aforementioned, it is essential to take measures to improve the security of maritime communications. As such, the authors in~\cite{Guo2022} presented analyses on cross-layer attacks and security measures in the satellite MCNs. In addition, the authors in~\cite{Jiang2023} explored a secure and robust communication scheme by modeling the channel phase and angle uncertainties of geostationary Earth orbit and LEO. However, the encryption and decryption methods in the abovementioned works have limitations, as transmitting large amounts of data requires significant computational energy, resulting in transmission delays.

\par In this case, PLS can dynamically adapt security mechanisms according to the channel conditions, achieving secure transmissions for the maritime networks~\cite{Qian2022}. For example, the authors in~\cite{Xiong2024} aimed to enhance the performance of the satellite-terrestrial MCN by optimizing the transmission beamforming of the base station and LEO satellites. However, for fast-moving LEO satellites, the immediate processing and regulating of beamforming parameters demand significant resources, increasing computational load. As such, low-altitude platforms are employed to introduce friendly jamming signals, which can significantly enhance PLS~\cite{Huang2025}. {\color{black}Common low-altitude platforms, such as eVTOL aircrafts and UAVs, play key roles in various applications~\cite{Huang2024b}. Moreover, UAVs, with low cost, high mobility, and flexible deployment, are well-suited for implementing maritime PLS~\cite{Qian2022}. For example, the authors in~\cite{Wang2019a} considered a power allocation scheme to optimize secrecy throughput to improve the PLS of downlink transmission. In addition, the authors in~\cite{Liu2022} presented a reinforcement learning-enabled UAV maritime communication relay strategy with a dueling structure to resist jamming attacks. However, the aforementioned works overlook the complex trade-offs between conflicting objectives such as security performance and energy efficiency, as well as the varying importance of these objectives across different scenarios, which often leads to suboptimal solutions.}

\subsection{Optimization Approaches}

\par To address these challenges, MOP provides a mathematical framework for optimizing multiple conflicting objectives simultaneously~\cite{Karami2022}. By formulating an MOP, we can model the correlations between conflicting objectives and find solutions that offer a better compromise across varying conditions. Generally, several common methods exist for handling MOP. First, swarm intelligence optimization algorithms can solve MOP, which gradually approximates the Pareto front and obtains multiple non-dominated solutions~\cite{Zheng2024}. {\color{black}For example, the authors in~\cite{Huang2025} proposed a collaborative beamforming method to resist eavesdropping via UAVs, and presented a swarm intelligence algorithm to address MOP. Moreover, the authors in~\cite{Sun2023} considered a UAV-assisted communication scenario where eavesdroppers aim to intercept the data, and proposed a multi-objective salp swarm algorithm to deal with the MOP. However, swarm intelligence algorithms lack dynamic adjustment and timely feedback mechanisms, which affects overall performance and is not applicable to real-time problems.
}

\begin{table}[!t]
\renewcommand{\arraystretch}{1.1}
\newcommand{\tabincell}[2]{\begin{tabular}{@{}#1@{}}#2\end{tabular}}
\caption{Main notations}
\label{table:notations}
\centering
\begin{tabular}{ll}
\hline
\bfseries Notation & \bfseries Definition\\
\hline
 & Notation in the system model\\
\hline
$\omega$ & Argument of periapsis \\
$\varsigma$ &  Rician fading\\
$\rho $ & Air density\\
$a_{r}$ & Rotor disk area\\
$d_{U,V}$ & Distance between the UAV and vessel\\
$d_{S,V}$ & Distance between the LEO satellite and vessel\\
$F_s$ &  Rician factor \\
$h_{U,V}$  & Channel of the UAV to vessel\\
$h_{S,V}$ & Channel of the LEO Satellite to vessel \\
$I_0$ & Maximum value of allowable interference power\\
$P_{B}$ & Blade profile power\\
$P_{I}$ & Induced power\\ 
$PL_{U,V}$ & Path loss between the UAV and vessel\\
$PL_{S,V}$ & Path loss between the LEO satellite and vessel\\
$r_{a}$ & Airframe drag ratio \\
$s_{r}$ & Rotor robustness\\
$v_{f}$ & Forward direction velocity of the UAV\\
$v_{i}$ & Average rotor induced velocity\\
$v_{h}$ & Horizontal direction velocity of the UAV\\
$v_{tip}$ & Rotor blade tip velocity\\
$v_{v}$ & Vertical direction velocity of the UAV\\
\hline
& Notation in the algorithm\\
\hline
$\tau_{m}$ & Weights for the optimization objective $m$ \\
$\alpha$ & Temperature parameter of SAC\\
$\mathcal{D}$ & Replay buffer \\ 
$\mathcal{A}$ & Action set\\
$J_{m}(\pi)$ & The expected return for the optimization objective $m$\\ 
$\boldsymbol{K}$ & Key matrix \\
$N(a)$ & Number of arms of MAB \\
$\boldsymbol{Q}$ & Query matrix \\
$Q_{m}^{\theta}$ & Q estimated value for the optimization objective $m$ \\
$\mathcal{R}_{m}$ & Reward for the optimization objective $m$ \\
$\mathcal{S}$ & State set\\
$\boldsymbol{V}$ & Value matrix \\
$V_{m}^{\psi}$ & State-value for the optimization objective $m$\\
$V_{m}^{\hat{\psi}}$ & Target state-value for the optimization objective $m$\\
\hline
\end{tabular}
\end{table}

\par Furthermore, DRL has adaptive learning capabilities and can handle complex action spaces, making it suitable for solving dynamic timing problems~\cite{Li2024}. For example, the authors in~\cite{Wang2021} considered a dual-UAV secure maritime communication system that maximizes the minimum secrecy rate for the mobile user over all time slots. Moreover, the authors in~\cite{Yang2024} considered a UAV reconfigurable intelligent surface-assisted maritime communication system to maximize energy efficiency and ensure the quality of requirements against jamming attacks. {\color{black}However, the aforementioned works overlook crucial challenges such as the local optimal problem in the long-term optimization process, computational burden in large-scale action spaces, and suboptimal solutions due to improper weight value settings.

\par In summary, different from previous works, we employ low-altitude UAV friendly-jamming for LEO satellite-maritime communications and propose a novel GenAI-enabled DRL approach that can derive a high-quality policy for resolving dynamic SSMCMOP. This approach is more suitable for dynamic scenarios requiring trade-offs between conflicting objectives while enabling fast response to changing conditions.}

%
\section{Models and preliminaries} \label{sec:models_and_formulataion}

\par {\color{black}In this section, we first present the LEO satellite-maritime communication system assisted by a low-altitude UAV.} Then, we introduce the LEO satellite orbit model and vessel movement model. Next, the corresponding communication network is presented. Finally, the energy consumption model of the UAV is presented. In addition, the main notations are shown in Table~\ref{table:notations}.

\subsection{System Overview}

\par
\begin{figure}
\centering
\includegraphics[width=3.5in]{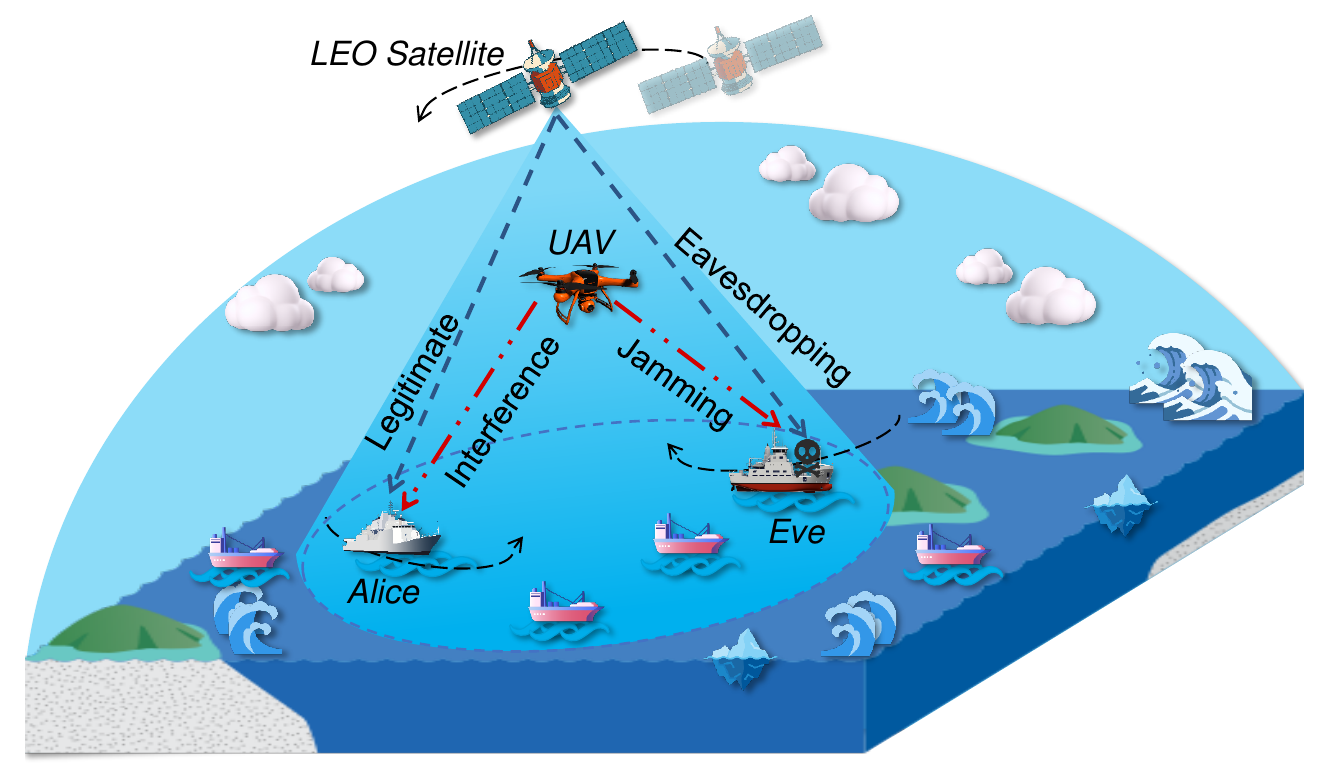}
\caption{{\color{black}An LEO satellite-maritime communication system assisted by low-altitude UAV.}}
\label{fig:scenario-model}
\end{figure}

 {\color{black}As shown in Fig. \ref{fig:scenario-model}, we consider an LEO satellite-maritime communication system assisted by a low-altitude UAV, which involves an LEO satellite, a UAV, a legitimate vessel denoted as Alice, and an illegitimate user denoted as Eve.} Among them, the LEO satellite receives information from the data fusion center and forwards it to the target vessel Bob within range. Note that the LEO satellite is equipped with high-performance antennas with adequate power, which can ensure efficient downlink communication to the vessels. However, its open channel is vulnerable to eavesdropping by Eve. In this case, due to the dynamics of the marine channels and constraints of marine routes, it is challenging to combat the eavesdropping attack with existing vessels. Moreover, the offshore evaporation ducts and multipath scattering increase the complexity of vessel-aided methods. Therefore, we mainly focus on the low-altitude friendly-jamming platforms to deal with eavesdropping attacks. Among the platforms, UAVs, with great mobility, high flexibility, and wide coverage, are particularly suited for implementing secure maritime communications~\cite{Liu2022a}.

\par Without loss of generality, we take into account a discrete-time system that operates in a finite time $\mathcal{T}$, $\mathcal{T} = \left \{ 1,2,\dots, T \right \}$. The LEO satellite follows its fixed orbit, and the vessels (Alice and Eve) navigate on their predetermined trajectories. In this case, the LEO satellite sends signals to Alice over a legitimate link, whereas Eve aims to eavesdrop on the data content by the eavesdropping link. To enhance the reliability of the LEO satellite communications, the UAV sends jamming signals to Eve. Moreover, the UAV is configured with a single omni-directional antenna and optical camera, which can sense and detect the position of the eavesdropping vessel when the vessel temporarily regulates position and direction.

\par During the process of satellite-maritime communications, we utilize the 3D Cartesian coordinate system to indicate the time-varying positions of Alice, Eve, the UAV, and the LEO satellite at time slot $t$ as ($x_{A}[t]$, $y_{A}[t]$, $z_{A}[t]$), ($x_{E}[t]$, $y_{E}[t]$, $z_{E}[t]$), ($x_{U}[t]$, $y_{U}[t]$, $z_{U}[t]$), and ($x_{S}[t]$, $y_{S}[t]$, $z_{S}[t]$), respectively. The satellite spectrum resources are limited and UAV needs to share spectrum with satellites, which complicates the dynamic communications~\cite{Li2020}. To this end, we model the LEO satellite orbit and the vessel movement, as well as the communication processes of the LEO satellite and vessels, the UAV and vessels, to express the dynamics of the system.

\begin{figure}
\begin{minipage}[t]{0.25\textwidth}
\centering
\includegraphics[width=1.8in]{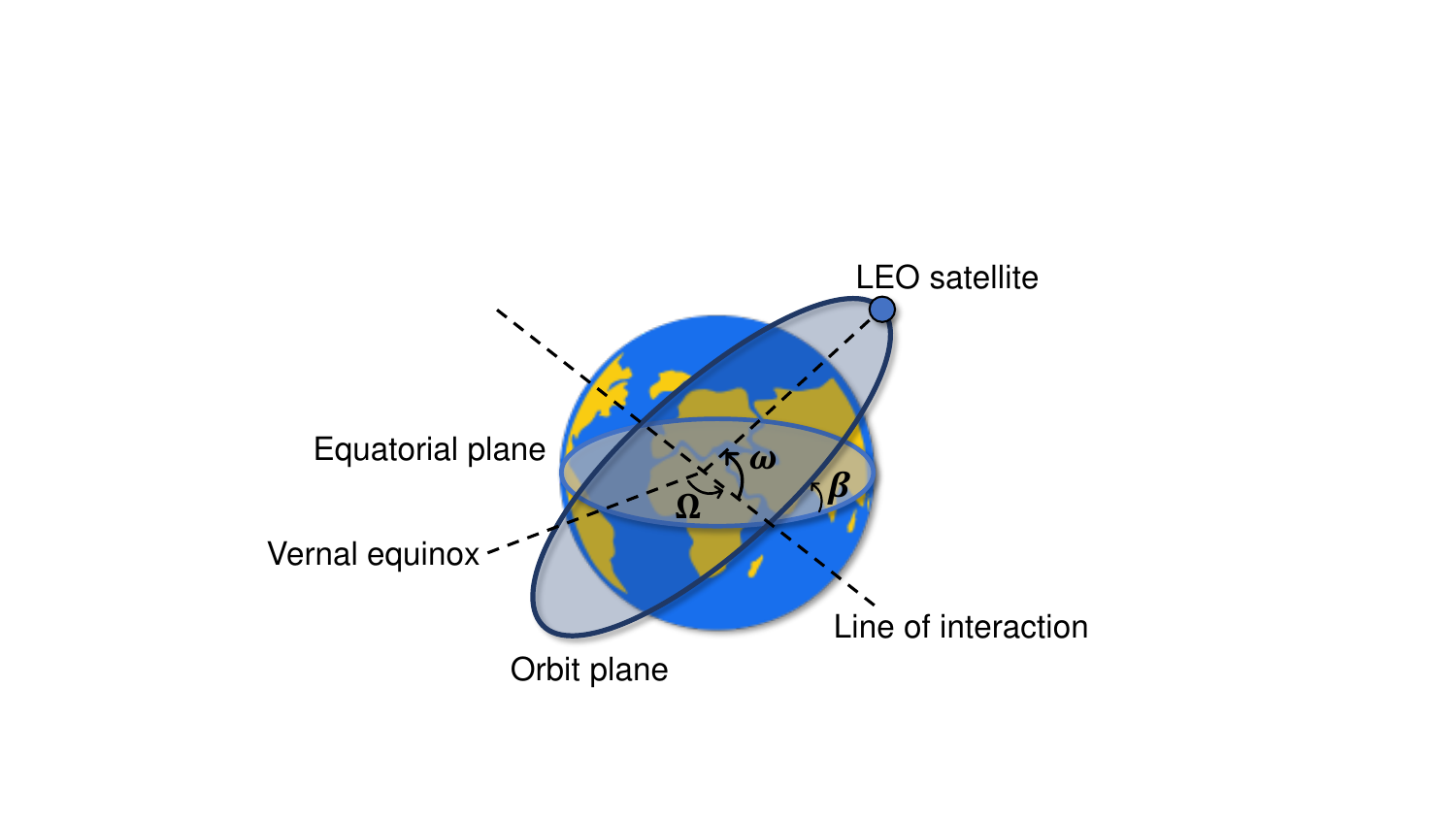}
\caption{LEO satellite orbit model.}
\label{fig:satellite}
\end{minipage}%
\begin{minipage}[t]{0.25\textwidth}
\centering
\includegraphics[width=1.6in]{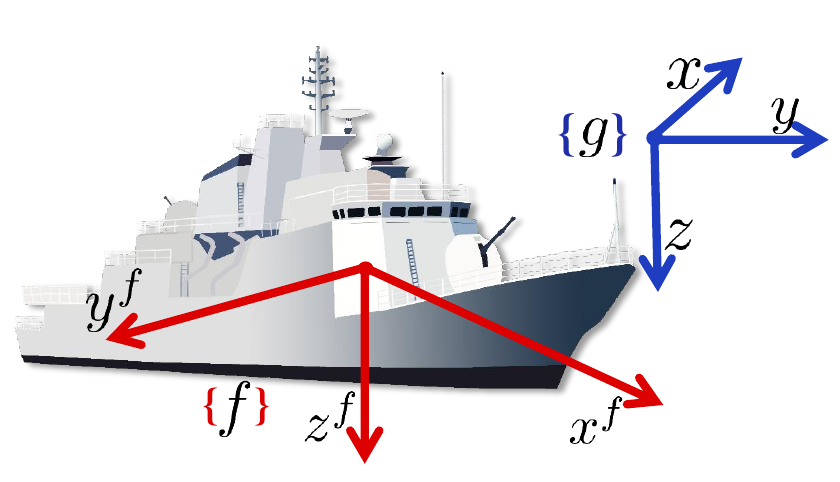}
\caption{Vessel coordinate systems.}
\label{fig:vessel}
\end{minipage}
\end{figure}

\subsection{LEO Satellite Orbit Model}

\par LEO satellites orbit at altitudes ranging from about 500 to 2,000 kilometers above the surface of Earth~\cite{Deng2021}. Since the orbital altitudes are relatively low, LEO satellites are characterized by fast orbital speeds, usually circling the Earth every 90 to 120 minutes~\cite{Pokhrel2024}. Mathematically, the motion of the LEO satellite is usually described by the Keplerian six elements $\left \langle \beta, \omega, \Omega, e, a, \vartheta \right \rangle$~\cite{Deng2021}, as shown in the orbit model in Fig. \ref{fig:satellite}, which is introduced in detail as follows.
\begin{itemize}
    \item \textit{Inclination Angle ($\beta $):} It is the intersection angle between the orbital plane of the LEO satellite and the equatorial plane. Specifically, the satellite is moving in the opposite direction of rotation of the Earth if $\beta$ is more than 90$^{\circ}$.
    \item  \textit{Argument of Periapsis ($\omega$):} It denotes the angle between the direction of the LEO satellite and intersections of the orbital and equatorial planes.
    \item \textit{Right Ascension of Ascending Node ($\Omega$):} It denotes the angle between the vernal equinox and intersections of the orbital and equatorial planes.
    \item \textit{Eccentricity ($e$):} It is the eccentricity of the orbital ellipse.
    \item \textit{Semi-Major Axis ($a$):} It is the distance from the center of the track to the furthest point on the edge of the track.
    \item \textit{True Anomaly ($\vartheta$):} It is the angle between the satellite direction and perigee direction.
\end{itemize}

\par We consider that the LEO satellite orbit is circular~\cite{Deng2021}. In this case, $e$ and $\vartheta$ are set to 0, and $a$ is equal to the orbital radius $l_{S}$. Accordingly, the elements of the LEO satellite in $m$ orbit at $t$ time can be set as $\left \langle \beta _m, \omega_m[t], \Omega_m, {l}_{Sm} \right \rangle$, wherein $\omega_m[t]=\omega_m[t]+2\pi \left ( ( {t}/{\mathcal{T}_{Sm}} ) \mod 1\right ) $ and $\mathcal{T}_{Sm}$ denotes the orbital period of the LEO satellite. Moreover, $l_{Sm}=H_{Sm}+R_{E}$, where $H_{Sm}$ and $R_E$ are the orbital altitude of the LEO satellite and the radius of Earth, respectively. In this case, the orientation of the LEO satellite in the 3D Cartesian coordinates $(x_{Sm}[t], y_{Sm}[t], z_{Sm}[t])$ at time slot $t$ is expressed by
\begin{equation}
\begin{split}
   & \left(\begin{array}{l}
x_{Sm}[t] \\
y_{Sm}[t] \\
z_{Sm}[t]
\end{array}\right)=\\&
l_{Sm}\left(\begin{array}{c}
\cos \omega_m[t] \cos \Omega_m-\sin \omega_m[t] \cos \beta_m \sin \Omega_m \\
\cos \omega_m[t] \sin \Omega_m+\sin \omega_m[t] \cos \beta_m \cos \Omega_m \\
\sin \omega_m[t] \sin \beta_m
\end{array}\right).
\end{split}
\label{LEO position}
\end{equation}

\subsection{Vessel Movement Model}

\par As shown in Fig. \ref{fig:vessel}, two three-dimensional right-handed Cartesian coordinate systems of a vessel are generally used to represent movement of the vessel~\cite{Skulstad2020}. Specifically, one is a general coordinate system $\left \{ g \right \} $ in which the origin is on the surface of the sea, and the defined $x$, $y$, and $z$ axes are pointing the north, east, and down, respectively. The other is a fixed coordinate system $\left \{ f \right \} $ in which the origin is the gravity of the vessel, with $x^{f}$, $y^{f}$, and $z^{f}$ pointing to the bow, starboard, and down, respectively. The rotations around the $x^{f}$, $y^{f}$, and $z^{f}$ axes are defined as roll ($\phi$), pitch ($\theta$), and yaw ($\psi$), respectively. Moreover, the Euler angle vector is denoted as $\boldsymbol{\Theta}=\left [ \phi, \theta, \psi \right ] ^{T}$. Mathematically, the six degree-of-freedom (DOF) vessel model $\left \langle x,y,z, \phi, \theta, \psi \right \rangle $ is used to represent movement of a vessel, which is as follows: 
\begin{equation}
    \dot{ \boldsymbol{\eta}}[t]=\boldsymbol{\Gamma} (\boldsymbol{\Theta}[t]) \boldsymbol{\nu}[t],
    \label{vessel-position1}
\end{equation}
\noindent where $ \boldsymbol{\eta}[t]=\left [ x[t],y[t],z[t],\phi[t], \theta[t],\psi[t] \right ] ^{T}$ indicates the displacement and rotation vector at time slot $t$, and $ \boldsymbol{\nu}[t]=\left [ u[t],v[t],w[t],p[t],q[t],r[t] \right ] ^{T}$ is the vector of translational and rotational velocities at time slot $t$. Moreover, $\dot{\boldsymbol{\eta}}[t]$ denotes the first derivative of $\boldsymbol{\eta}$, and $\boldsymbol{\Gamma}$ denotes the horizontal plane rotation matrix from $\left \{ f \right \} $ to $\left \{ g \right \} $. Moreover, the velocity vector is often related to the corresponding forces caused by the wind, waves, propulsion, and inertia factors, and it is calculated by
\begin{equation}
\begin{split}
\left(\boldsymbol{M}_{R}+\boldsymbol{M}_{A}\right) \dot{\boldsymbol{\nu}}[t]+&\boldsymbol{C}(\boldsymbol{\nu}[t]) \boldsymbol{\nu}[t] + \boldsymbol{D}(\boldsymbol{\nu}[t]) \boldsymbol{\nu}+\boldsymbol{g}(\boldsymbol{\eta})\\ &=\boldsymbol{\tau}_{\text {th}}[t]+\boldsymbol{\tau}_{\text {wind }}+\boldsymbol{\tau}_{\text {cur }}+\boldsymbol{\tau}_{\text {wave}},
\end{split}
\label{vessel-position2}
\end{equation}
\noindent where $\boldsymbol{M}_{R}$, $\boldsymbol{M}_{A}$, $\boldsymbol{C}(\boldsymbol{\nu})$, and $\boldsymbol{D}(\boldsymbol{\nu})$ denote the matrices of the rigid-body mass, added mass, Coriolis, and damping coefficient, respectively. Moreover, $\dot{\boldsymbol{\nu}}[t]$ denotes the first derivative of $\boldsymbol{\nu}$, $\boldsymbol{g}(\boldsymbol{\eta})$ denotes the resilience, $\boldsymbol{\tau}_{\text {th}}[t]$ is the vessel thrusters vector at time slot $t$, $\boldsymbol{\tau}_{\text {wind}}$, $\boldsymbol{\tau}_{\text {cur}}$, and $\boldsymbol{\tau}_{\text {wave}}$ indicate vectors of force on the vessel arising from wind, current, and wave, respectively.

\par The relative positions of the vessel and LEO satellite affect the satellite-vessel signal transmission, and the relative positions of the vessel and the UAV determine the quality of jamming signal transmission. Next, we describe the corresponding communication model in detail.

\subsection{Communication Model}

\par In the designed system, our concerned communication links include the LEO satellite-to-vessel (S2V) link and UAV-to-vessel (U2V) link. Specifically, the S2V link is employed for transmitting data signals, which are at risk of being eavesdropped by Eve. Moreover, the U2V link is utilized to jam Eve, which may interfere with the effective data reception of Alice. The two communication links are elaborated in detail below.

\subsubsection{S2V Link from LEO Satellite}

\par During the S2V link, we consider that the vessels can detect the signals from the LEO satellite by using the approach in~\cite{Guo2022}, thus obtaining the quantized form of the actual channel state information (CSI). In this case, since the altitude of the LEO satellite is sufficient for line-of-sight (LoS) transmission~\cite{Shang2024}, we employ a typical composite channel to represent the communications from the LEO satellite to vessels. The channel of the S2V link at time slot $t$ can be expressed by~\cite{Li2020}
\begin{equation}
    h_{{S}, {V}}[t]=\sqrt{PL_{{S}, {V}}[t]} \left ( \sqrt{\frac{F_{S}}{1+F_{S}}}+\sqrt{\frac{1}{1+F_{S}}} M_{S,V}[t]\right ),
    \label{channel satellite to vessel}
\end{equation}
\noindent where $F_S$ denotes the Rician factor, and $M_{S,V}[t]\in \mathcal{CN} (0,1)$. Moreover, the path loss $PL_{{S}, {V}}[t]$ between the LEO satellite and vessel is defined by
\begin{equation}
    PL_{{S}, {V}}[t] (dB) =C_{S}+10 W_{S} \log 10\left(d_{S,V}[t]\right)+\delta _{S,V}[t],
\label{pathloss S2V}
\end{equation}
\noindent where $C_{S}$ and $W_{S}$ are the path loss parameter and exponent, respectively. Moreover, ${d_{S, V}[t]}$ is the distance between the LEO satellite and vessel (Alice or Eve) at time slot $t$, which is computed by the 3D positions of the LEO satellite and vessel according to Eqs. (\ref{LEO position}) and (\ref{vessel-position1}), respectively. In addition, $\delta_{S,V}[t]$ is the zero-mean Gaussian random variable with standard deviation $\sigma_{X_S}$~\cite{Wang2018},~\cite{Wu2018a}. Note that $d_{S, A}[t]$ and $d_{S, E}[t]$ are the distance from the LEO satellite to Alice and Eve at time slot $t$, respectively, which are used to calculate corresponding path loss ($PL_{S, A}[t]$ and $PL_{S, E}[t]$), and get the matching channels ($h_{S, A}[t]$ and $h_{S, E}[t]$).

\subsubsection{U2V Link from UAV}

\par In the considered system, the antenna on the UAV is notably higher than that of the vessel. Thus, the path loss of jamming signals between the UAV and the vessel at time slot $t$ is calculated as follows:
\begin{equation}
    {PL_{U,V}[t]} (dB) = C_{U} + 10 W_{U}\log10 \left(\frac{d_{U,V}[t]}{d_{c}}\right)+\delta _{U,V}[t],
    \label{PL_{U,V}}
\end{equation}
\noindent where ${d_{U, V}[t]}$ denotes the distance from the UAV to vessel (Alice or Eve) at time slot $t$, which is computed based on the movements of the UAV and vessel. Moreover, $C_{U}$ denotes the path loss parameter, $d_c$ denotes the reference distance, $W_{U}$ and $\delta _{U, V}[t]$ are the path loss parameter and zero-mean Gaussian random variable with $\sigma_{X_U}$, respectively~\cite{Matolak2016}. Note that $PL_{U, A}[t]$ and $PL_{U, E}[t]$ denote the path loss from the UAV to Alice and Eve at time slot $t$, respectively, which are obtained by the distance from the UAV to Alice and Eve ($d_{U, A}[t]$ and $d_{U, E}[t]$), respectively.

\par On the basis of the path loss between the UAV and vessel, the U2V link channel at time slot $t$ is indicated by
\begin{equation}
    h_{U,V}[t] = \frac{{\varsigma [t]}^2}{PL_{U,V}[t]}, 
\end{equation}
\noindent where $\varsigma [t] $ is the Rician fading at time slot $t$. Moreover, $h_{U, A}[t]$ and $h_{U, E}[t]$ are used to represent the channels from the UAV to Alice and Eve at time slot $t$, respectively.

\par According to the S2V link and U2V link, the achieved transmission rate of Alice at time slot $t$ is denoted by
\begin{equation}
    R_{{A}}[t]=\log _{2}\left(1+\frac{P_{{S}} G_{{S}} G_{{S,S}}\left|h_{{S}, {A}}[t]\right|^{2}}{P_{{U}}[t]G_{U} h_{U,A}[t]+\sigma^{2}}\right),
    \label{Alice_rate}
\end{equation}
\noindent where $P_{{S}}$ is the transmit power of the LEO satellite, $G_S$ and $G_{U}$ denote the antenna gains of the satellite and UAV, respectively. Moreover, $G_{S, S}$ is the antenna gain of the vessel served by the LEO satellite, $P_{{U}}[t]$ denotes the transmit power of the UAV at time slot $t$, and $\sigma^{2}$ denotes the maritime additive white Gaussian noise power.

\par Furthermore, the reachable transmission rate of Eve at time slot $t$ is denoted by
\begin{equation}
    R_{{E}}[t]=\log_{2}\left(1+\frac{P_{{S}} G_{{S}} G_{{S,S}}\left|h_{{S}, {E}}[t]\right|^{2}}{P_{{U}}[t]G_{U} G_{U,E} h_{U,E}[t]+\sigma^{2}}\right),
    \label{Eve_rate}
\end{equation}
\noindent where $G_{U, E}$ denotes the antenna gain of Eve served by the UAV, as the UAV acts as a jammer mainly towards the eavesdropper Eve.

\par Thereby, with $R_{A}[t]$ and $R_{E}[t]$, the secrecy rate from the LEO satellite to Alice at time slot $t$ is expressed by
\begin{equation}
     R_{SEC}[t] = \left [ R_{A}[t]-R_{E}[t] \right ]^+,
     \label{secrecy rate}
\end{equation}
\noindent where $\left [ \chi \right ]^+$ indicates the value at which the larger of $0$ and $\chi$.

\par From the aforementioned discussion, the 3D position and transmit power of the UAV are critical parameters for regulating the communication network. Note that the position of the UAV changes continuously over $\mathcal{T}$ time slots, resulting in inevitable energy consumption. Therefore, the following section details the UAV energy consumption model. 

\subsection{UAV Energy Consumption Model}

\par We consider that in each time slot, the UAV performs action $\mathbf{a}^{U}[t] =(a_{x}^{U}[t], a_{y}^{U}[t], a_{z}^{U}[t])$ to move. Thereby, at time slot $t$, the positional regulation of the UAV can be determined by the action $\mathbf{a}^{U}[t]$, which can be expressed as $(x_{U}[t],y_{U}[t],z_{U}[t]) = (x_{U}[t-1],y_{U}[t-1],z_{U}[t-1]) + \mathbf{a}^{U}[t]$.

\par Furthermore, we introduce the moving UAV energy consumption. Commonly, the total energy consumption of a UAV is composed of communication energy and propulsion energy. However, the communication energy is usually ignored during the calculation because its value is extremely small compared to propulsion energy~\cite{Zeng2018}. Thus, the propulsion power consumption of the UAV in 2D horizontal state is calculated by~\cite{Zeng2018}
\begin{equation}
\begin{split}
    P_p(v_{h}[t]) &= P_{I}\left [ \sqrt{1+\left ( \frac{v_{h}^{2}[t] }{2v_{i}^{2}} \right )^2 } -\left (\frac{v_{h}[t]}{\sqrt{2}v_{i}} \right )^2 \right ]^{\frac{1}{2}} + \\ &P_{B}\left [ 1+\left ( \frac{\sqrt{3} v_{h}[t]}{v_{tip}} \right ) ^2\right ] +\frac{1}{2}r_{a}s_{r}a_{r}\rho v_{h}[t]^{3},
\end{split}
\label{p_v}
\end{equation}
\noindent where $v_{h}[t] = \sqrt{(a_{x}^{U}[t])^{2}+(a_{y}^{U}[t])^{2}} /\Delta t$ is the horizontal direction velocity of the UAV, $P_{I}$ and $P_{B}$ are the induced power and blade profile power, respectively, $v_{tip}$ and $v_{i}$ are the rotor blade tip velocity and average rotor induced velocity, respectively. Moreover, $r_{a}$, $s_{r}$, $a_{r}$, and $\rho$ denote the airframe drag ratio, rotor robustness, rotor disk area, and air density, respectively.

\par Note that we exclude the energy consumed by the UAV accelerating or decelerating, since this process occupies only a minor fraction of the total UAV flight time~\cite{Li2023a}. Therefore, we use a heuristic closed approximation to express the 3D energy consumption of the UAV, where the considerations include the propulsive energy, kinetic energy, and gravitational energy during the ascent and descent of the UAV over time. The 3D trajectory energy consumption of the UAV is denoted by
\begin{equation}
\begin{split}
E_{U}(T)\approx\int_{0}^{T} &P_p\left(v_{h}[t]\right) dt+\frac{1}{2}m_{U}\left (v_{f}[T]^{2}-v[0]^{2} \right)+\\&m_{U}g\left(h[T]-h[0]\right ),
\end{split}
\label{E_t}
\end{equation}
\noindent where $v_{f}[t] = \sqrt{\left ( v_{h}[t]\right )^{2} + \left ( v_{v}[t]\right )^{2} }$ is the UAV forward direction velocity at time slot $t$, of which $v_{v}[t] = |a_{z}^{U}[t]|/\Delta t$ denotes the vertical direction velocity of the UAV. Moreover, $h[0]$ and $h[T]$ are the UAV flight altitudes at start time and end time, respectively, $m_{U}$ and $g$ denote the mass and gravitational acceleration of the UAV, respectively.

%
\section{Problem formulation and analyses} \label{sec:problem formulation and analysis}

\par In this section, we first state the optimization problem, and then analyze the problem.

\subsection{Problem Statement}

\par In the considered scenario, the trajectories of the LEO satellites and vessels are not controllable. Specifically, the vessels sail along specific routes dictated by engines and perform tasks such as sensing and data collection. Moreover, the LEO satellites operate in particular orbits, and their movements are governed by orbital mechanics. The altitude and inclination of the orbit of the satellite are precisely designed to achieve a specific observation or communication mission. 

\par In this case, this work utilizes the UAV to transmit jamming noise to Eve to resist eavesdropping, thereby enabling secure communications from the LEO satellite to Alice. Due to the spectrum scarcity problem, the UAV and the LEO satellite share the spectrum~\cite{Li2020}. The jamming link between the UAV and Eve may interfere with Alice. To address this, we need to regulate the jamming signals transmitted by the UAV, improving its effect on Eve while minimizing the interference on Alice. Therefore, we focus on maximizing the secrecy rate as shown in Eq. (\ref{secrecy rate}), which can be controlled by the \textit{3D position} and \textit{transmit power} of the UAV. Note that regulations in the 3D position inevitably result in additional energy consumption of the UAV. Therefore, minimizing the \textit{positional regulations} of the UAV is essential to enhance overall energy efficiency.

\par Combining the aforementioned factors, the decision variables to be jointly optimized are the following UAV-related parameters: \textit{(i)} $\mathbb{L} $ = \{$\mathbb{X}, \mathbb{Y}, \mathbb{Z}$\} denotes the 3D location set of the UAV over $\mathcal{T} $ time slots, where $\mathbb{X}$ = $ \left \{ x_{U}[t]\right \}_{t=0}^{T}$, $\mathbb{Y}$ = $ \left \{  y_{U}[t]\right \}_{t=0}^{T}$, and $\mathbb{Z}$ = $ \left \{  z_{U}[t]\right \}_{t=0}^{T}$. \textit{(ii)} $\mathbb{P}$ = $ \left \{ P_{U}[t]\right \}_{t=0}^{T}$ is the transmit power of the UAV over $\mathcal{T}$ time slots.

\subsection{Problem Formulation}

\par In the LEO satellite-maritime communication system assisted by a low-altitude UAV, we contemplate the following two optimization objectives.

\par \textbf{\textit{Optimization Objective 1:}} To achieve secure LEO satellite-maritime communications, the first optimization objective is to maximize the \textbf{\textit{average secrecy rate}} over $\mathcal{T}$ time slots by regulating the 3D position and transmit power of the UAV, which is expressed by
\begin{equation}
    f_{1}(\mathbb{L},\mathbb{P})=\frac{1}{T} {\textstyle \sum_{t=0}^{T}} R_{SEC}[t].
\end{equation}

\par \textbf{\textit{Optimization Objective 2:}} In the process of accomplishing the first optimization objective, the UAV needs to constantly adjust the position, increasing energy consumption. Given the limited energy supply at sea, the second optimization objective is to minimize the \textbf{\textit{average energy consumption}} of the UAV over $\mathcal{T}$ time slots, which is expressed by
\begin{equation}
f_{2}(\mathbb{L})=\frac{1}{T} {\textstyle \sum_{t=0}^{T}} E_{U}[t],
\end{equation}
\noindent where $E_{U}[t]$ denotes the energy consumption of the UAV at time slot $t$.

\par {\color{black}Maximizing the average secrecy rate requires regulating the position of the UAV, which conflicts with minimizing the average energy consumption of the UAV. Moreover, according to Eq. (\ref{p_v}), higher UAV speeds lead to increased energy consumption. Conversely, as the UAV slows down, communication time increases, leading to higher hovering energy consumption. Consequently, the two optimization objectives are in conflict, requiring a suitable modeling method to balance this conflicting relationship. In this case, the multi-objective optimization problem modeling provides a mathematical framework that simultaneously optimizes multiple conflicting objectives, which is well-suited for capturing trade-offs of conflicting metrics and can be used to formulate our optimization problem~\cite{Karami2022}.}
\par According to the aforementioned optimization objectives, we formulate the SSMCMOP as follows:
\begin{subequations}
\label{con:mop}
\begin{align}
\min _{\left \{ \mathbb{L},\mathbb{P} \right \} } \  &F=\left \{ -f_{1},f_{2} \right \}, \label{con:mopZa}\\
\mbox{s.t.}\  &C1: x_{min}\le x_{U}[t] \le x_{max}, \forall t\in  \mathcal{T},\label{con:mopZb}\\   
&C2: y_{min}\le y_{U}[t] \le y_{max}, \forall t\in  \mathcal{T},\label{con:mopZc}\\ 
&C3: z_{min}\le z_{U}[t] \le z_{max}, \forall t\in  \mathcal{T},\label{con:mopZd} \\
&C4: P_{min}\le P_{U}[t] \le P_{max}, \forall t\in  \mathcal{T},\label{con:mopZe} \\
&C5: {\textstyle \sum_{t=0}^{T}} P_{U}[t]  \Delta t \leq E_{0}, \forall t\in  \mathcal{T},\label{con:mopZf} \\
&C6: P_{{U}}[t]G_{U} h_{U,A}[t] \le I_{0},  \forall t\in  \mathcal{T},\label{con:mopZg} \\
&C7: P_{{U}}[t]G_{U}G_{U,E} h_{U,E}[t] \le I_{0},\forall t\in  \mathcal{T},\label{con:mopZh}
\end{align}
\end{subequations}
\noindent where $C1$ and $C2$ indicate the horizontal flight ranges of the UAV, $C3$ denotes the vertical flight height of the UAV, and $C4$ indicates the transmit power limitation of the UAV. Moreover, $C5$ denotes the energy consumption constraint, where $E_{0}$ is the maximum allowable energy consumption of the UAV over $\mathcal{T}$ time slots. In addition, $C6$ and $C7$ are the interference temperature limitations of the UAV, where $I_{0}$ is the maximum allowable interference power to ensure that the interference does not excessively affect the communications of other legitimate devices.

\subsection{Problem Analyses}

\par Subsequently, we present the corresponding analyses of the formulated SSMCMOP.

\par \textit{{\color{black}(i) The SSMCMOP is a dynamic and large-scale problem:}} In our considered scenario, both the LEO satellites and vessels move along the respective orbits, reflecting a realistic setup. In this case, the UAV needs to transmit jamming signals based on the position of the vessel, leading to a dynamic communication channel. At this point, the SSMCMOP involves two optimization objectives over $\mathcal{T}$ time slots, both of which change in real-time. {\color{black}In addition, the UAV has multiple variables (3D position and transmit power) to be optimized. Therefore, SSMCMOP is a dynamic and large-scale problem.}

\par \textit{(ii) The SSMCMOP is with long-term optimization objectives:} The changes of the LEO satellite, vessels, and UAV can lead to fluctuations in signal strength, which in turn affects the optimization objectives. Moreover, since we consider optimization objective values over $\mathcal{T}$ time slots, an optimal solution at any specific moment may not necessarily represent the optimal solution over a longer time scale. Therefore, SSMCMOP involves long-term optimization objectives and requires balancing current objectives with long-term objectives.

\par \textit{(iii) The SSMCMOP is an NP-hard problem:} To simplify the analysis, we only focus on the first optimization objective. Specifically, we fix the positions of vessels and the LEO satellite and set $P_{U}[t]$ to discrete. In this case, the simplified SSMCMOP is expressed by
\begin{subequations}
\label{con:NP-prove}
\begin{align}
\min _{\left\{ \mathbb{L},\mathbb{P} \right \} } \ &F= -f_{1} , \label{con:npa}\\
\mbox{s.t.}\ & \text{Eqs.} (\ref{con:mopZb})- (\ref{con:mopZd}), (\ref{con:mopZf})- (\ref{con:mopZh}),\\
& P_{U}[t] \in \{0, P_{max}\}, \forall t\in \mathcal{T}, \label{con:npc}\\
&{\textstyle \sum_{t=0}^{T}} P_{U}[t] < T P_{max}, \forall t\in \mathcal{T}.\label{con:npd}
\end{align}
\end{subequations}

\par Clearly, the simplified SSMCMOP is a classic nonlinear multidimensional 0-1 knapsack problem, which has been proved to be NP-hard~\cite{Goos2020}. Consequently, the original continuous SSMCMOP is an NP-hard problem.

\par In summary, since the SSMCMOP presents significant challenges, traditional convex optimization methods and evolutionary computation methods struggle to address the dynamic problem~\cite{Wang2023}. {\color{black}In this case, the DRL algorithms can adaptively learn strategies through continuous environmental interaction in the online training phase, followed by offline execution where the trained model can quickly generate actions to respond to real-time changes~\cite{Li2024}. Therefore, we employ a DRL-based algorithm to tackle the SSMCMOP.}

%
\section{Algorithm} 
\label{sec:algorithm}

\par In this section, we first transform the SSMCMOP into an MDP, then introduce the process of conventional SAC. Next, considering the weaknesses of SAC for MDP, we propose a TransSAC algorithm to address these challenges. 

\subsection{MDP Formulation}

\par In the considered scenario, our main concern is ensuring the availability of trained DRL models. To this end, we transform the SSMCMOP into an MDP. Specifically, MDP is a mathematical framework used to model decision-making in an uncertain environment, defined by $\left \{ \mathcal{S, A, P},\mathcal{\mathbf{R}}, \gamma \right \} $~\cite{Xu2020}, which is introduced in detail as follows.

\par Each state $\boldsymbol{s}[t] \in \mathcal{S}$ describes the situation of the system at time slot $t$, given the current state $\boldsymbol{s}[t] $, and the agent can choose the action $\boldsymbol{a}[t]\in \mathcal{A}$. Moreover, $\mathcal{P} \left ( \boldsymbol{s}[t+1]| \boldsymbol{s}[t], \boldsymbol{a}[t] \right ) $ is the probability of reaching to state $\boldsymbol{s}[t+1]$ after taking action $\boldsymbol{a}[t]$. Then, the reward function $\mathcal{\mathbf{R}} =\left \{ \mathcal{R}_{1}{\left ( \boldsymbol{s}[t], \boldsymbol{a}[t] \right )}, \mathcal{R}_{2}{\left ( \boldsymbol{s}[t], \boldsymbol{a}[t] \right )} \right \} $ is used to evaluate the effectiveness of the decision by calculating the immediate reward for $\boldsymbol{a}[t]$ according to the optimization objectives. In addition, the discount factor $\gamma \in \left [ 0, 1 \right) $ is utilized to weigh the relative importance of current and future rewards. MDP aims to determine a policy $\pi$ that maximizes the expected cumulative reward. For the optimization objective $m$, the expected return is defined by
\begin{equation}
    J_{m}(\pi)=\frac{1}{T}\mathbf{E}\{ {\textstyle \sum_{t=0}^{T}}  \gamma \mathcal{R}_{m}(\boldsymbol{s}[t], \boldsymbol{a}[t]) | \boldsymbol{s}[0] =\boldsymbol{s}, \boldsymbol{a}[0]=\boldsymbol{a}\},
\end{equation}
\noindent where $\mathbf{E}\{ \cdot \} $ is the expectation according to the policy $\pi$. Moreover, the state space, action space, and reward function of SSMCMOP are introduced in detail as follows.

\subsubsection{State Space} At each time slot, the agent focuses on relevant state information to make decisions. In this system, the agent (i.e., the UAV) is concerned with its own relevant information, and the positions of the LEO satellite and vessels. Since the UAV is equipped with a positioning device, and the locations of the vessels and LEO satellites can be computed based on their models given in Section~\ref{sec:models_and_formulataion}, obtaining the data is feasible. As such, the state of the system is expressed by
\begin{equation}
\begin{split}
&\mathcal{S}= \{ \boldsymbol{s}[t] | \boldsymbol{s}[t]= (\boldsymbol{\eta}_{A}[t], \boldsymbol{\nu}_{A}[t],\boldsymbol{\eta}_{E}[t], \boldsymbol{\nu}_{E}[t], P_{U}[t], \\ & ( x_{U}[t], y_{U}[t], z_{U}[t]), (x_{Sm}[t], y_{Sm}[t], z_{Sm}[t]) ) ,\forall t\in \mathcal{T} \},
\end{split}
\end{equation}
\noindent where $\boldsymbol{\eta}_{A}[t]$ and $\boldsymbol{\nu}_{A}[t]$ are the position and velocity of Alice at time slot $t$, respectively. Moreover, $\boldsymbol{\eta}_{E}[t]$ and $\boldsymbol{\nu}_{E}[t]$ are the position and velocity of Eve at time slot $t$, respectively. {\color{black} Note that the current scenario includes a UAV with a single antenna, which limits the UAV to jam one illegitimate vessel at a time. Furthermore, our algorithm can be extended to scenarios with multiple UAVs, and the details are provided in Section \ref{sec:Discussion}.}

\subsubsection{Action Space} In the considered system, the UAV acts as an agent that can take actions based on the current state. Thus, the action space includes the 3D position and transmit power of the UAV, which is denoted by
\begin{equation}
    \mathcal{A}= \{\boldsymbol{a}[t]| \boldsymbol{a}[t]= (\mathbf{a}^{U}[t],P_{U}[t]),\forall t \in \mathcal{T} \}. 
\end{equation}

\subsubsection{Immediate Reward} 

\par {\color{black}The reward mechanism acts as a feedback signal to direct the agent in a series of actions and directly affects the quality of the final policy. Therefore, designing an effective reward method is critical to improve overall performance. {\color{black} Note that the constraints $C1$-$C5$ of the SSMCMOP are satisfied by setting the parameters of the UAV within the predetermined ranges. Moreover, the constraints $C6$ and $C7$ are satisfied by incorporating them into the reward function as penalty items, and we further incorporate constraints $C1$, $C2$, and $C3$ into the reward function to ensure strict adherence to flight boundaries.} In our work, the MDP employs a reward function in vector form, which is defined by
\begin{equation}
\begin{split}
\mathcal{\mathbf{R}}&= \left\{\mathcal{R}_{1}(\boldsymbol{s}[t], \boldsymbol{a}[t]), \mathcal{R}_{2}(\boldsymbol{s}[t], \boldsymbol{a}[t])\right\} \\& = \left\{\begin{array}{ll}
\left(\mu_{1}W_{c}[t] R_{SEC}[t], -\mu_{2}W_{c}[t] E_{U}[t]\right), & k[t]=1,\\
\left(\mu_{1}\varrho_{1}W_{c}[t] R_{SEC}[t], -\mu_{2} \varrho_{2}W_{c}[t] E_{U}[t]\right), &k[t]=0,
\end{array}\right.
\end{split}
\end{equation}
\noindent where $\mathcal{R}_{1}[t]$ and $\mathcal{R}_{2}[t]$ denote the scaled reward values of the optimization objectives $R_{SEC}[t]$ and $E_{U}[t]$, respectively. Moreover, when the UAV moves within the defined range at time slot $t$, $k[t]$ is assigned to 1. Otherwise, $k[t]$ is assigned to 0. Moreover, the coefficients $\mu_{1}$ and $\mu_{2}$ are proportionality factors used to ensure that the rewards for both targets are on the same order of magnitude. Additionally, $\varrho_{1}$ and $\varrho_{2}$ are used to penalize when the UAV leaves the service area. Additionally, $W_c[t] = W_1[t] W_2[t]$ represents the comprehensive penalty term of the constraints $C6$ and $C7$ of the SSMCMOP. Correspondingly, $W_1$ and $W_2$ represent the penalty items when the strength of the interference signals received by Alice and Eve affects the maritime communications, respectively.}

\par {\color{black}Following this, we utilize the linear weighting approach to compute the overall expected return~\cite{Marler2010}. This approach offers low computational complexity and adjustable priority, making it well-suited for the SSMCMOP in resource-constrained dynamic maritime environments. The overall expected return is denoted by}
\begin{equation}
    J(\pi)=\tau_{1} J_{1}(\pi)+\tau_{2} J_{2}(\pi),
\end{equation}
\noindent where $\tau_{1}$ and $\tau_{2}$ are the weights of the two optimization objectives, with $\tau_{1} + \tau_{2} = 1$.

\subsection{Conventional SAC} 

\par Next, we discuss the advantages of SAC in dealing with MDP and describe the process in detail.

\subsubsection{Advantages of SAC}

\par General DRL approaches, such as the deep Q-network (DQN), are typically effective for discrete action space~\cite{fan2020theoretical}, whereas discretizing for continuous action space problems is not feasible. Moreover, trust region policy optimization (TRPO) can enhance policy stability by optimizing the trust region, while the computational complexity is higher~\cite{Schulman2015}. In contrast, SAC is well-suited for continuous-time problems~\cite{Zhang2021}. \textit{First}, by incorporating maximum entropy theory, SAC promotes policy diversity, enabling better adaptation and exploration in complex environments and avoiding local optima. \textit{Second}, the offline data update strategy further enhances sample efficiency through the iterative use of the replay buffer. \textit{Finally}, SAC integrates the policy gradient approach with the Q-value function for policy updates, which improves both sample efficiency and training stability. Therefore, we select SAC as the framework for addressing the MDP.

\subsubsection{Process of SAC}

\par A crucial feature of SAC is the maximum entropy theory, which enhances policy randomness and improves the exploratory capability. The expected return, including the entropy term, is expressed by
\begin{equation}
\begin{split}
    J_{m}(\pi)=\frac{1}{T}\mathbf{E} \{ {\textstyle \sum_{t=0}^{T}} \gamma  \left[  \mathcal{R}_{m} \left ( \boldsymbol{s}[t], \boldsymbol{a}[t]\right ) - \alpha \log \pi ( \boldsymbol{a}[t]|\boldsymbol{s}[t]) \right] \},
\end{split}
\end{equation}
\noindent where $\log \pi (\boldsymbol{a}[t]| \boldsymbol{s}[t])$ denotes the entropy of the policy $\pi$, which promotes explorability. Moreover, $\alpha$ is the temperature parameter to balance the reward and entropy~\cite{Zhang2021}.

\par Furthermore, the value $Q^{\theta}$ can evaluate the soft Q-value network which is parameterized by $\theta$. To reduce the correlation between the input data, the outcome of each step is saved in the replay buffer $\mathcal{D}$, and the network performance can be evaluated by using the mean square error (MSE) as follows:
\begin{equation}
L_{m}(\theta )=\mathbf{E} \{ \frac{1}{2} [ Q_{m}^{\theta} \left ( \boldsymbol{s}[t], \boldsymbol{a}[t]\right )- \hat{Q}_{m}\left ( \boldsymbol{s}[t], \boldsymbol{a}[t]\right ) ]^{2} | \mathcal{D}  \},
\label{loss-Q}
\end{equation}
\noindent where $Q_{m}^{\theta}(\boldsymbol{s}[t], \boldsymbol{a}[t])$ and $\hat{Q}_{m}(\boldsymbol{s}[t], \boldsymbol{a}[t])$ are the Q estimated value and Q target value for the optimization objective $m$ at time slot $t$, respectively, and $\hat{Q}_{m}(\boldsymbol{s}[t], \boldsymbol{a}[t])$ is computed by
\begin{equation}
\hat{Q}_{m}\left (\boldsymbol{s}[t], \boldsymbol{a}[t]\right ) = \mathcal{R}_{m} \left ( \boldsymbol{s}[t], \boldsymbol{a}[t]\right ) + \gamma \mathbf{E}\{V_{m}^{\hat{\psi} }(\boldsymbol{s}[t+1]) \},
\label{target state value}
\end{equation}
\noindent where ${V}_{m}^{\hat{\psi} }(\boldsymbol{s}[t+1])$ is the target state-value, which will be introduced later. To minimize the $L_{m}(\theta)$, stochastic gradient $\nabla_{\theta}L_{m}(\theta)$ is commonly used to optimize the parameter $\theta$.

\par To improve training stability, the state-value network ${V}_{m}^{\psi}$ parameterized by $\psi$ is introduced. Similarly, we use MSE to evaluate network performance as follows: 
\begin{equation}
\begin{split}
    L_{m}&(\psi)=\mathbf{E}\{\frac{1}{2}(V_{m}^\psi (\boldsymbol{s}[t])-\\ &\mathbf{E}\{Q_{m}^{\theta}\left(\boldsymbol{s}[t], \boldsymbol{a}[t]\right)-\alpha \log {\pi_{\Phi}}\left(\boldsymbol{a}[t]|\boldsymbol{s}[t]\right)|\pi_{\Phi }\})^{2}|\mathcal{D} \},
    \end{split}
\label{state value}
\end{equation}
\noindent where ${V}_{m}^{\psi}(\boldsymbol{s}[t])$ denotes the state-value for the optimization objective $m$ at time slot $t$, and $\psi$ is updated by the stochastic gradient $\nabla_{\psi}L_{m}(\psi)$. Moreover, to stabilize the critic network update, the parameter $\hat{\psi}$ can estimate ${V}_{m}^{\hat{\psi} }(\boldsymbol{s}[t])$ by the soft update. Furthermore, $\Phi$ indicates the policy network parameter, and the corresponding policy loss function is denoted by
\begin{equation}
\begin{split}
     L(\Phi)=\mathbf{E}\{&\alpha\log \pi_{\Phi}\left(f_{\Phi}(\epsilon[t] ; \boldsymbol{s}[t]) |\boldsymbol{s}[t]\right)- \\ &{\textstyle \sum_{m=1}^{2}}\tau_{m}Q_{m}^{\theta}\left(\boldsymbol{s}[t], f_{\Phi}(\epsilon[t] ; \boldsymbol{s}[t]) \right) | \mathcal{D}, \mathcal{N}\},
    \label{loss-pi}
\end{split}
\end{equation}
\noindent where $ f_{\Phi}(\epsilon[t] ; \boldsymbol{s}[t]) $ denotes a reparameterization trick, and $\epsilon$ is the action noise sampled from a stationary distribution $\mathcal{N}$. Accordingly, $\Phi$ is updated by gradient descent $\nabla_{\Phi}L(\Phi)$.

\subsection{The Proposed TransSAC Algorithm}

\subsubsection{Motivation of TransSAC Algorithm}

\par While SAC has advantages in solving continuous-time problems, it still faces the following challenges.

\begin{figure*}[!t]
\centering
\includegraphics[width=7in]{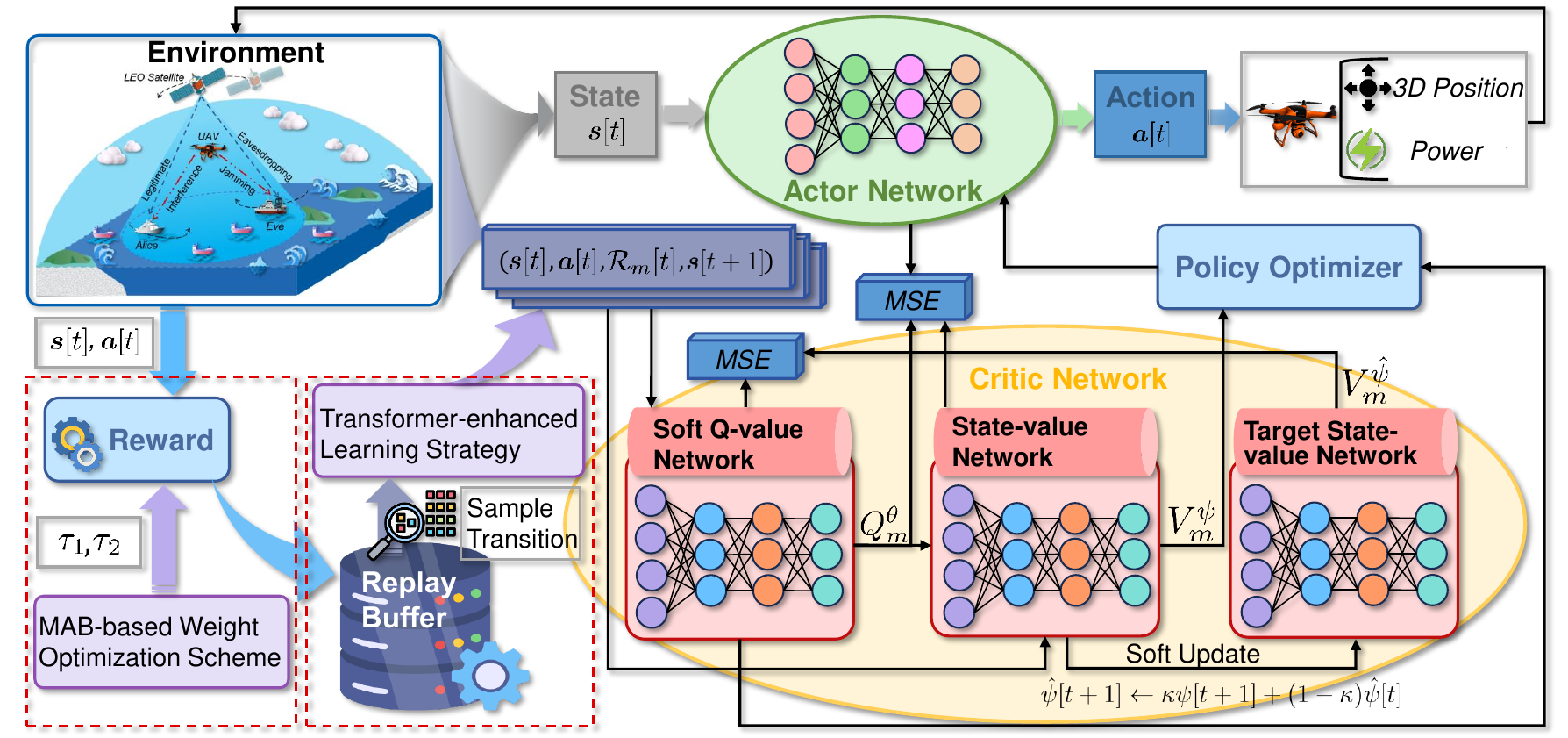}
\caption{The framework of the proposed TransSAC algorithm for solving the SSMCMOP, where a transformer-enhanced learning strategy and an MAB-based weight optimization scheme are integrated into the network to capture global dependencies and explore weights diversely.}
\label{fig:sac}
\end{figure*}

{\color{black}
\par \textit{(i) Strong Temporal Correlation:} The vessels and LEO satellites are governed by constrained moving trajectories, and the UAV flight paths need to satisfy physical continuity constraints. These constraints make their current positions strictly limited by previous states, thereby causing the MDP to exhibit strong temporal correlation. At this point, conventional SAC primarily focuses on the current state and ignores the strong temporal correlation, which may result in ineffective decision-making. Note that this may cause policies to fall into local optima and further affect the performance of the optimization objectives.

\par \textit{(ii) Large-Scale State and Action Spaces:} In the MDP, the UAV needs to process large state inputs and select action spaces. This results in the MDP having large-scale state and action spaces where multiple action dimensions need to be jointly optimized, resulting in combinatorial complexity. However, conventional SAC relies on trial-and-error learning to discover relevant patterns and optimize policies. As such, large-scale state and action spaces result in SAC requiring more training rounds to interact with the environment to discover effective policies, which reduces sampling efficiency. 
}
\par \textit{(iii) Preset Suboptimal Weights for Multiple Optimization Objectives:} In the MDP, we aim to simultaneously optimize multiple objectives. Generally, conventional SAC uses fixed weights for each optimization objective, yet sensitive weight configurations may make it potentially difficult to adapt to the long-term dynamic problem. Moreover, improper weight configurations may lead to suboptimal performance. At this point, SAC may prioritize energy consumption over security, and vice versa. This makes it challenging to effectively balance the conflicting objectives when the preset weights are suboptimal.

\par Therefore, considering the aforementioned weaknesses of the conventional SAC, we propose a TransSAC algorithm, and Fig. \ref{fig:sac} shows the framework of the proposed TransSAC algorithm for solving the SSMCMOP. Moreover, the overall structure is illustrated in Algorithm \ref{alg:algorithm1}, and the corresponding improvements are introduced as follows.

\subsubsection{Transformer-enhanced Learning Strategy}

\par {\color{black} The transformer, a key GenAI approach, employs the self-attention mechanism to compute the global correlation weights across temporal samples, so that it can establish comprehensive feature interactions between multiple states~\cite{Han2021}. As such, by incorporating the transformer into standard DRL algorithms, we can enhance temporal modeling capabilities, allowing policy networks to explicitly capture long-term dependencies and sequential patterns. This process effectively addresses biases in locally correlated samples, thereby overcoming the challenge of local optima in conventional SAC. Furthermore, the transformer employs parallelized sequence processing to efficiently handle large-scale state and action spaces. Moreover, through multi-head attention mechanisms, the transformer can decompose high-dimensional actions and process different dimensions separately, improving the ability of the neural network to fit policy functions. Therefore, we use the transformer to process actions and states to improve the performance of the conventional SAC. Note that this approach does not conflict with the Markov property of the MDP, as the approach does not rely on complete historical information. Instead, the approach can enhance the representation of the current state and action, providing contextual information to optimize decision-making. The overall framework is presented in Algorithm \ref{alg:algorithm2}, with the details as follows.
}

\par \textit{First}, positional encoding is a key component used to give positional information for each element in the sequence~\cite{Vaswani2017}. In the MDP, positional encoding provides relative or absolute temporal positions for the inputs (states and actions), which can be denoted by~\cite{Vaswani2017}
\begin{subequations}
\label{con:trans2}
\begin{align}
&\mathcal{E}_{2i} = \sin(\frac{P[t]}{\varpi ^{2i/d_{e}}} ),\label{con:trans2-a}\\
&\mathcal{E}_{2i+1} = \cos(\frac{P[t]}{\varpi ^{2i/d_{e}}}), \label{con:trans2-b}
\end{align}
\end{subequations}
\noindent where $\mathcal{E}_{2i}$ and $\mathcal{E}_{2i+1}$ are the positional encoding of even and odd numbers, respectively. Moreover, $P[t]$ indicates the position of the inputs at time slot $t$, $\varpi$ is the related constant, and $d_{e}$ is the embedding dimension of the model.

\par {\color{black}\textit{Second}, the self-attention mechanism is a key for the transformer, which can capture global information by quantifying the relevance between each action and current state. This allows algorithms to efficiently evaluate the subsequent impact of decisions, thereby solving the challenge of local optima caused by strong temporal correlation.} Accordingly, the self-attention mechanism is expressed by~\cite{Vaswani2017}
\begin{equation}
    \mathcal{M}_{SA}(\boldsymbol{Q}, \boldsymbol{K}, \boldsymbol{V})=\mathcal{U}_{softmax}\left(\frac{\boldsymbol{Q} \boldsymbol{K}^{T}}{\sqrt{d_{\boldsymbol{K}}}}\right) \boldsymbol{V},
    \label{attention1}
\end{equation}
\noindent where $\boldsymbol{Q}=\boldsymbol{H} \boldsymbol{W}^{Q}$, $\boldsymbol{K}=\boldsymbol{H} \boldsymbol{W}^{K}$, and $\boldsymbol{V}=\boldsymbol{H} \boldsymbol{W}^{V}$ denote the query, key, and value matrices, respectively, with $\boldsymbol{W}^{Q}, \boldsymbol{W}^{K}$, and $ \boldsymbol{W}^{V}$ indicate the corresponding learnable weight matrices, and $\boldsymbol{H}$ denotes the joint denotation of $\mathcal{S}$ and $\mathcal{A}$. Moreover, $\mathcal{U}_{softmax}$ is the normalization operation performed using the softmax function, and $d_{\boldsymbol{K}}$ denotes the dimension of the $\boldsymbol{K}$-matrix. {\color{black}Note that the parallelized sequence processing capability of the self-attention mechanism can comprehensively analyze entire state-action sequences, effectively avoiding random and ineffective explorations and greatly improving sample efficiency in the large action space.}

\begin{algorithm}
\label{alg:algorithm1}
\caption{TransSAC Algorithm}
\KwIn {Number of iterations, batch size, smoothing parameter $\kappa$, and learning rates.} 
Initialize the parameters with soft Q-value network $\theta$, policy network $\Phi$, state-value network $\psi$, and target state-value network $\hat{\psi}$, and initialize replay buffer $\mathcal{D}$;\\
\For{each iteration}{
\For{each environment step}{
Obtain weights by \textbf{Algorithm \ref{alg:algorithm3}};\\
Select and execute action $\boldsymbol{a}[t]$, $\boldsymbol{a}[t]\sim \pi_{\Phi}(\boldsymbol{a}[t]|\boldsymbol{s}[t])$;\\
Observe next state $\boldsymbol{s}[t+1]$ and reward $\boldsymbol{R}$;\\
Update replay buffer $\mathcal{D}$, $\mathcal{D} \gets \mathcal{D} \cup \left ( \boldsymbol{s}[t], \boldsymbol{a}[t],\mathcal{R}_{m}(\boldsymbol{s}[t], \boldsymbol{a}[t]), \boldsymbol{s}[t+1] \right ) $;\\
Obtain enhanced states and actions by \textbf{Algorithm \ref{alg:algorithm2}};\\
}
\For{each gradient step}{
Calculate the MSE of the state-value network by Eq. (\ref{state value}) and update parameter $\psi$;\\
Soft update the target state-value network;\\
Calculate the Q target value $\hat{Q}_{m}$ by Eq. (\ref{target state value});\\
Compute the MSE of the soft Q-value network by Eq. (\ref{loss-Q}) and update parameter $\theta$;\\
Calculate weighted policy network loss by Eq. (\ref{loss-pi}) and update parameter $\Phi$;\\
}
}
\KwOut{Trained model.}
\end{algorithm}

\par {\color{black}\textit{Third}, multi-head attention mechanism decomposes high-dimensional actions into subspaces, enabling different attention heads to focus on different action dimensions. This approach can improve the fitting ability of the neural network,} which is defined by
\begin{subequations}
\label{con:trans4}
\begin{align}
   & \mathcal{M}_{MSA}(\boldsymbol{Q}, \boldsymbol{K}, \boldsymbol{V})= \mathcal{C}\left(\mathcal{H}_{1}, \ldots, \mathcal{H}_{h}\right) \boldsymbol{W}^{O},\label{con:trans4-a}\\
&\mathcal{H}_{i}=\mathcal{M}_{SA}\left(\boldsymbol{H} \boldsymbol{W}_{i}^{Q}, \boldsymbol{H} \boldsymbol{W}_{i}^{K}, \boldsymbol{H} \boldsymbol{W}_{i}^{V}\right), \label{con:trans4-b}
\end{align}
\end{subequations}
\noindent where $\mathcal{H}_{i}$ is the output of attention head $i$, $\mathcal{C}$ is the concat processing, and $\boldsymbol{W}^O$ is the output weight matrix. Moreover, the attention output for each $\mathcal{H}$ is computed independently and then linearly transformed through $\boldsymbol{W}^O$.

\par \textit{Finally}, the output after the self-attention layer is fed into a feed-forward neural (FFN), which performs an independent nonlinear transformation of the output at each time slot. As such, the FFN enhances the capability of the model to learn complex patterns, and the transformation is given by~\cite{Vaswani2017}
\begin{equation}
\label{FFN}
    \mathcal{M}_{FFN}(x)= \left [ x\boldsymbol{W}_1+\boldsymbol{b}_1  \right ] ^+ \boldsymbol{W}_2 +\boldsymbol{b}_2,
\end{equation}
\noindent where $x$ denotes the input vector, and it is also the output obtained from the previous layer. Moreover, $\boldsymbol{W}_{1}$ and $\boldsymbol{W}_2$ denote the weight matrices of the first and second linear transformations, respectively, $\boldsymbol{b}_{1}$ and $\boldsymbol{b}_{2}$ are corresponding bias vectors.

\par {\color{black}In summary, the transformer-enhanced learning strategy can effectively address the challenge of strong temporal correlation through enhanced state and action representations while handling large-scale state and action spaces through parallelized processing. These capabilities make the transformer particularly well-suited for our large-scale and long-term optimization problem.}

\begin{algorithm}
\label{alg:algorithm2}
\caption{Transformer-enhanced Learning Strategy}
\KwIn {States and actions from $\mathcal{D}$.}
Add positional encoding using Eqs. (\ref{con:trans2-a}) and (\ref{con:trans2-b});\\
Perform a linear transformation of $\boldsymbol{Q}$, $\boldsymbol{K}$, and $\boldsymbol{V}$;\\
Calculate the attention values for all heads with the softmax function by Eqs. (\ref{con:trans4-b}) and (\ref{attention1});\\
Splice the output of all heads by Eq. (\ref{con:trans4-a});\\
Apply residual linking and layer normalization;\\ 
Perform FFN with two linear layers by Eq. (\ref{FFN});\\
\KwOut{Enhanced states and actions representations.}
\end{algorithm}

\subsubsection{Multi-armed bandit (MAB)-based Weight Optimization Scheme}

\par Different from the preset weights, we employ an MAB-based weight optimization scheme that dynamically explores weights during the multi-objective optimization process. The MDP is dynamic with long-term optimization objectives, and balancing the conflicting objectives is crucial and challenging. Moreover, fixed and preset weights may prioritize one objective over others, causing algorithms to converge to a local optimum over time. In this case, the MAB-based scheme continuously and diversely explores weights, allowing the algorithm to discover a broader solution space. Specifically, by using the $\varepsilon$-greedy strategy, MAB balances exploration and exploitation, reducing the risk of falling into local optima. In addition, MAB does not require complex models or enormous computational resources, enabling quick execution. This makes it well-suited for real-time optimization within the system. The general process is presented in Algorithm \ref{alg:algorithm3}, and the details are introduced as follows.

\par In the MAB, each arm $a$ represents a weight, MAB selects an arm and calculates the corresponding reward for estimating the arm. The updated rule of reward is defined by
\begin{equation}
    \mathcal{R}_{n}(a)=\mathcal{R}_{n}(a)+ \frac{\mathcal{R}-\mathcal{R}_{o}(a)}{N(a)},
    \label{MAB1}
\end{equation}
\noindent where $\mathcal{R}_{o}(a)$ and $\mathcal{R}_{n}(a)$ are the old and new rewards, respectively. Moreover, $\mathcal{R}$ is the reward value for the time slot, and $N(a)$ indicates the number of times the arm has been chosen.

\par In summary, the MAB-based weight optimization scheme diversely explores weights for MDP, effectively balancing the solutions among conflicting optimization objectives. This avoids inappropriate weights that could lead to suboptimal performance and maintains computational efficiency.

\begin{algorithm}[t]
\label{alg:algorithm3}
\caption{MAB-based Weight Optimization Scheme}
Initialize the reward $\mathcal{R}_{n}(a)$ for each arm, selection count $N(a)$ for each arm, and probability $\varepsilon$;\\
Randomly initialize the weights $\tau_1$ and $\tau_2$;\\
\eIf{ random $< \varepsilon$}{
Randomly select an arm;
}
{
Select the arm with the maximum $\mathcal{R}_{n}(a)$;
}
Update $N(a)$ for the chosen arm;\\
Update $\mathcal{R}_{n}(a)$ for the chosen arm by Eq. (\ref{MAB1});\\ 
\KwOut{Optimal weights $\tau_1$ and $\tau_2$.}
\end{algorithm}

\subsection{Complexity Analysis of TransSAC Algorithm}

\par In this part, we analyze the computational complexity and space complexity of the TransSAC algorithm.

\par The overall computational complexity of the TransSAC algorithm is $\mathcal{O}(2|\theta|+|\Phi| + N_{t} M_{s} (|\Phi| + N(a)) + N_{t} M_{s} TD_{t} + N_{t} G (2|\theta|+|\Phi|))$, which is detailed as follows:
\begin{itemize}
    \item \textit{Network Initialization:} This process involves parameter initialization. The computational complexity is $ \mathcal{O}(2|\theta|+|\Phi|)$, where $|\theta|$ and $|\Phi|$ are the number of parameters in the critic and actor networks, respectively~\cite{Zhang2024a}.
    \item \textit{Action Selection:} Action selection is performed using the policy network. The computational complexity is $\mathcal{O} (N_{t} M_{s} (|\Phi| + N(a)))$, where $N_{t}$ indicates the number of training iteration, $M_{s}$ denotes the number of steps per iteration, and $N(a)$ denotes the arms in the MAB.
    \item \textit{Transformer Executing:} The complexity of executing the transformer is $\mathcal{O} (N_{t} M_{s} W_{t}D_{t})$, where $D_{t}$ indicates the data sampled from $\mathcal{D}$, and $W_{t}$ denotes the computational complexity of the transformer~\cite{Han2021}.
    \item \textit{Network Update:} For updating the critic and actor networks, the computational complexity is $\mathcal{O} (N_{t} G (2|\theta|+|\Phi|))$, where $G$ is the number of each gradient update.
\end{itemize}

\par The space complexity accounts for storing network parameters and the replay buffer, which contains states, actions, rewards, and next states tuples. Therefore, the space complexity of the TransSAC algorithm is $\mathcal{O}(2|\theta|+|\Phi|+N(a)+P_{T}+ D(2|\boldsymbol{s}|+|\boldsymbol{a}|+1))$, where $P_{T}$ is the storage space for transformer-related parameters, $D$ represents the replay buffer size, $|\boldsymbol{s}|$ and $|\boldsymbol{a}|$ are the dimensions of the state and action spaces, respectively.

%
\section{Simulation results and analyses} \label{sec:simulation-results-and-analysis}

\par In this section, we perform simulations to assess the performance of the TransSAC algorithm.

\subsection{Simulation Settings}

\subsubsection{Parameter Settings}

\par We conduct simulation experiments in Python 3.8 and Visual Studio Code 1.91 environments, and perform all the experiments on a server with AMD EPYC 7642 48-Core CPU, NVIDIA GeForce RTX 3090 GPU, and 128 GB RAM.

\par In the simulation, we randomly initialize the UAV within an 80 m × 80 m feasible flight region, as the UAV could be on another mission before. {\color{black} Note that our simulations account for vessel trajectory variability through stochastic environmental influences and dynamic random seed strategies, while varying the initial positions of LEO satellites across training episodes. These enable the agent to explore diverse states, enhancing simulation realism and facilitating comprehensive performance evaluation.} Moreover, for the TransSAC algorithm, each actor and critic network has two hidden layers and an output layer, with ReLU as the activation function, and the parameters are updated with the standard Adam optimizer. In addition, the batch size for sampling from the replay buffer is set to 128, and the number of attention heads and expansion multiplier of the hidden layer in the forward propagation are both set to 8. Additionally, the remaining primary parameters are presented in Table \ref{table：table1}~\cite{Deng2021},~\cite{Li2023a}.

\begin{table}[t]
\caption{{\color{black}Main parameters in the simulation process}}
\label{table：table1}
\begin{center} 
\renewcommand\arraystretch{1.1}
\begin{tabular}{lll}
\hline
\textbf{Notation} & \textbf{Definition} & \textbf{Value} \\
\hline 
$\beta_{m}$ & Inclination angle & 80$^{\circ}$\\
$\Omega_{m}$ & Right ascension of ascending node & 70$^{\circ}$ \\
$\sigma^{2}$ & Power of additive white Gaussian noise & -107 dBm\\
$\gamma$ & Discount factor & 0.9 \\
$ \kappa$ & Parameter for soft update & 0.005\\
$\varepsilon$ & Probability of exploration of MAB & 0.1\\
$\varpi$ & Constant of MAB & 10000\\
$C_{S}$ & Path loss parameter of the S2V link & 46.4\\
$C_{U}$ & Path loss parameter of the U2V link & 116.7 \\
$d_{c}$ & Reference distance of the U2V link & 2600 m\\
$E_{0}$ & Total communication energy of the UAV& 500 J\\
$F_{S}$ & Rician factor & 31.3 \\
$G_{E}$ & Gain of the vessel served by UAV & 8 dBi \\
$G_{S}$ & Antenna gain of the LEO satellite & 52 dBi\\
$G_{S,S}$ & Gain of the vessel served by satellite & 30 dBi\\
$G_{U}$ & Antenna gain of the UAV & 8 dBi \\ 
$H_{Sm}$ & Orbital altitude of the LEO satellite & 900 km\\
$I_0$ & Maximum allowable interference power & -74 dBm \\
$m_{U}$ & Aircraft mass & 2 kg\\
$P_{S}$ & Transmit power of the LEO satellite & 49.03 dBm\\
$R_{E}$ & Radius of the Earth & 6371 km\\
$T$ & Total time slots & 40 s \\
$W_{S}$ & Path loss constant exponent & 2\\
$W_{U}$ & Path loss constant exponent & 1.5\\
$z_{min}$ & Minimum altitude of the UAV & 50 m\\
$z_{max}$ & Maximum altitude of the UAV & 70 m\\
\hline
\end{tabular}
\end{center}
\end{table}

\subsubsection{Baselines}

\par {\color{black} To illustrate the performance of the TransSAC algorithm, we compare it with the theoretical optimal secrecy rate, a comparative approach, and various comparison algorithms as follows:}

\begin{figure*}
    \centering
       \subfloat[{LR.}]{\includegraphics[width=0.33\linewidth]{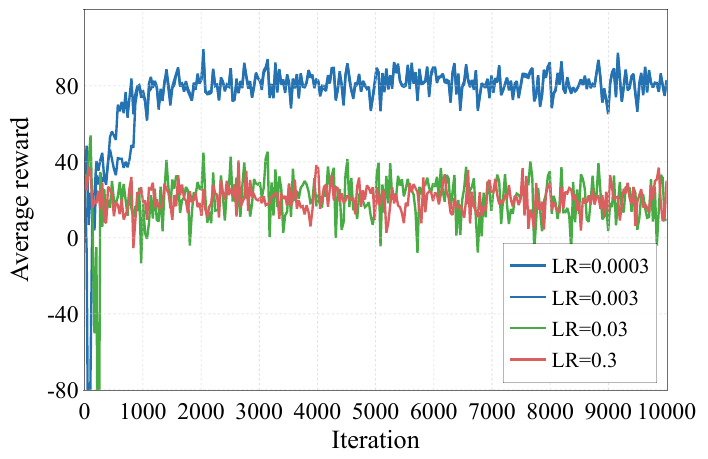}}
    \label{fig:lr}\hfill
   \subfloat[{Number of Neurons.}]{\includegraphics[width=0.33\linewidth]{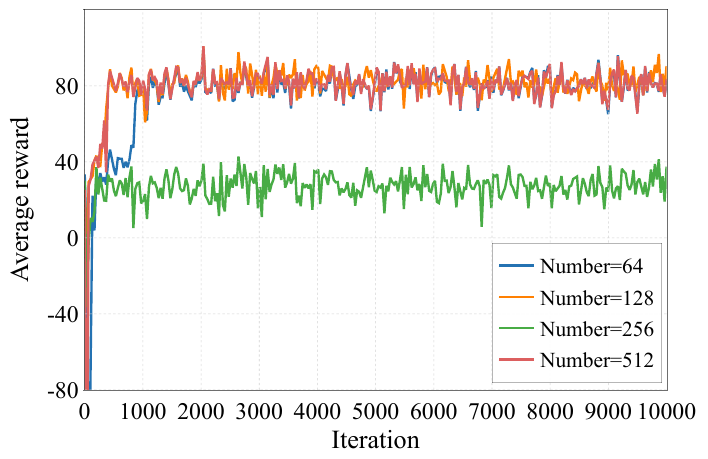}}
    \label{fig:number}\hfill
   \subfloat[{UR.}]{\includegraphics[width=0.33\linewidth]{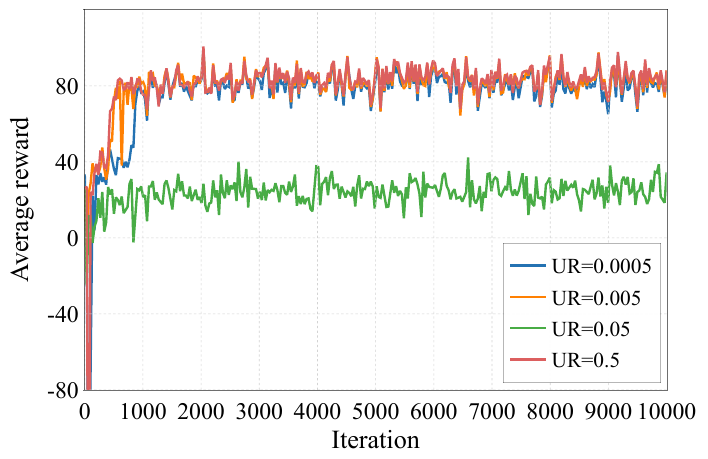}}
    \label{fig:Tau}\\
   \caption{Average reward of different hyper-parameters of the TransSAC algorithm.}
   \label{fig:compare2}
\end{figure*}

{\color{black}\textit{(i) Theoretical Optimal Secrecy Rate:} Under the fixed LEO satellite orbit and vessel moving trajectories, we consider a theoretical scenario where legitimate vessels receive LEO satellite signals without UAV interference, and the eavesdropper receives the maximum jamming power from the UAV. This idealized scenario allows us to calculate a theoretical optimal secrecy rate. Note that due to the NP-hard complexity and temporal characteristics of the problem, it is nearly impossible to obtain the theoretical optimal secrecy rate within real-world time constraints. Thus, this comparison indicates that our approach can obtain a near-optimal secrecy rate, making it more suitable and valuable for the considered scenario.}

\textit{(ii) Non-UAV Approach:} The approach relies on the LEO satellite sending signals to the legitimate vessel, without using UAV to interfere with the illegitimate vessel. As such, this comparison approach emphasizes the necessity of the UAV-assisted friendly-jamming approach in implementing secure LEO satellite-maritime communications.

\textit{(iii) State-of-the-art Comparison Algorithms:} We select deep deterministic policy gradient (DDPG)~\cite{Qiu2019}, twin delayed deep deterministic policy gradient (TD3)~\cite{Fujimoto2018}, proximal policy optimization (PPO)~\cite{Schulman2017}, and conventional SAC as comparison algorithms. These algorithms are commonly used to solve dynamic optimization problems~\cite{Zhang2024a}. Specifically, DDPG leverages the advantages of policy gradient approaches and deep learning, using the actor-critic structure to enhance policy learning, TD3 is a modification of the DDPG that improves performance through three aspects, including double Q-learning, delayed update, and target policy smoothing, and PPO improves the performance of the agent by optimizing the policy function and proposes trimming operations to maintain the stability and efficiency of training. The parameters of comparison algorithms are presented in Table \ref{table：table3}. Additionally, we set the total number of iterations in the aforementioned algorithms to $1 \times 10^6$, and evaluate these algorithms every 80 iterations during the training process.

\begin{table}[b]
\caption{Parameters of the comparison algorithms}
\label{table：table3}
\begin{center} 
\renewcommand\arraystretch{1.1}
\newcommand{\tabincell}[2]{\begin{tabular}{@{}#1@{}}#2\end{tabular}}
\begin{tabular}{ll}
\hline
\textbf{Algorithm} & \textbf{Parameters} \\
\hline 
DDPG& \tabincell{l}{$B_{z}$ = 128, $R_{U}$ = 0.005, $\theta$ = 0.15, $\sigma$ = 0.2,\\ $\gamma$ = 0.9, $R_{L}$ = 0.0003.}\\
\hline
PPO& \tabincell{l}{$B_{z}$ = 128, $R_{U}$ = 0.005, $\epsilon$ = 0.2, $\lambda$ = 0.9,\\ $\gamma$ = 0.9, $R_{L}$ = 0.0003, $C_{e}$ = 0.01.}\\
\hline
SAC& $B_{z}$ = 128, $R_{U}$ = 0.005, $\gamma$ = 0.9, $R_{L}$ = 0.0003. \\
\hline
TD3& \tabincell{l}{$B_{z}$ = 128, $F_{P}$ =2, $P_{nc}$ = 0.5, $P_{pn}$ = 0.2,\\$R_{U}$ = 0.005, $\gamma$ = 0.9, $R_{L}$ = 0.0003.}\\
\hline
\end{tabular}
\end{center}
\end{table}

\subsection{Simulation Results}

\subsubsection{Hyper-parameters Results} 

\par Since the hyper-parameters influence DRL performance, the key parameters of the TransSAC algorithm, including learning rate (LR), number of neurons, and update rate (UR), should be considered critically. Thus, we evaluate their impact on the TransSAC to determine optimal values.

\par \textit{(i) LR:} LR controls the step size of parameter updates, affecting training speed and stability. Fig. \ref{fig:compare2}(a) shows the learning process of the TransSAC under different LR values. As can be seen, the average rewards are highest and converge fastest when LR is set to 0.003. While an LR of 0.0003 achieves comparable rewards, the convergence speed is slower. Other LR settings show lower rewards, indicating suboptimal performance.

\par \textit{(ii) Number of Neurons:} The number of neurons in the hidden layer affects training efficiency and resource consumption. Fig. \ref{fig:compare2}(b) presents the performance of the TransSAC for different neurons. Specifically, the 128-neuron setting achieves the best performance, with higher rewards and minimal fluctuations after convergence. The 64-neuron setting performs reasonably while converging slower. In contrast, the 256 neurons result in lower rewards. Despite similar rewards to the 128-neuron setting, the 512-neuron setting suffers from higher variance, reducing stability.

\par \textit{(iii) UR:} UR controls the soft update speed, impacting training stability and convergence speed. Fig. \ref{fig:compare2}(c) describes the average reward curves of various UR values. As can be seen, a UR of 0.5 achieves the highest average rewards with the fastest convergence, making it the most effective setting. A UR of 0.005 yields comparable rewards yet converges slightly slower. A UR of 0.0005 stabilizes after convergence, while it is slower and less effective. In contrast, a UR of 0.05 results in much lower rewards, indicating suboptimal performance.

\begin{figure}
\centering
\includegraphics[width=3in]{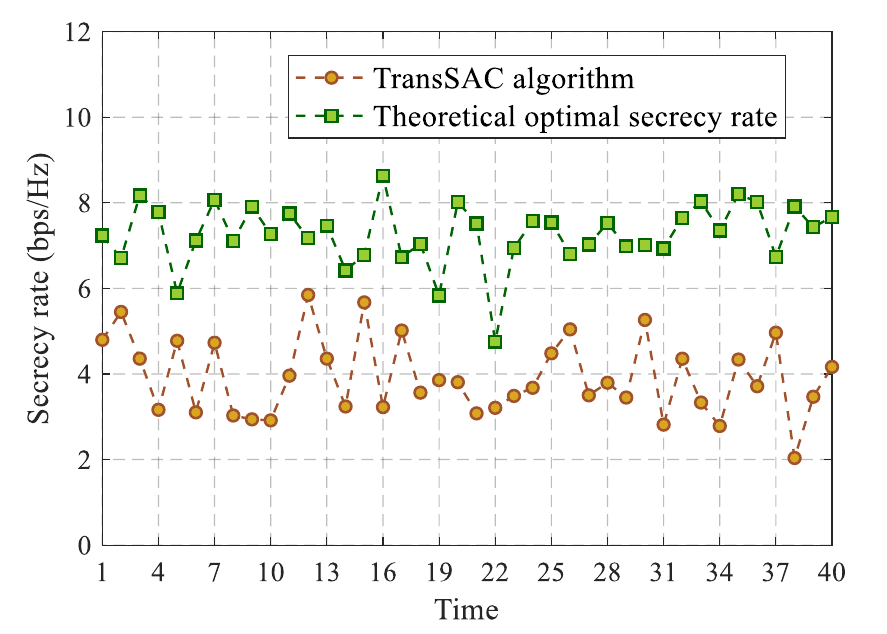}
\caption{Secrecy rate obtained by our approach in comparison to the theoretical optimal secrecy rate.}
\label{optimal_secracy_rate}
\end{figure}

\begin{figure}[b]
\centering
\includegraphics[width=3in]{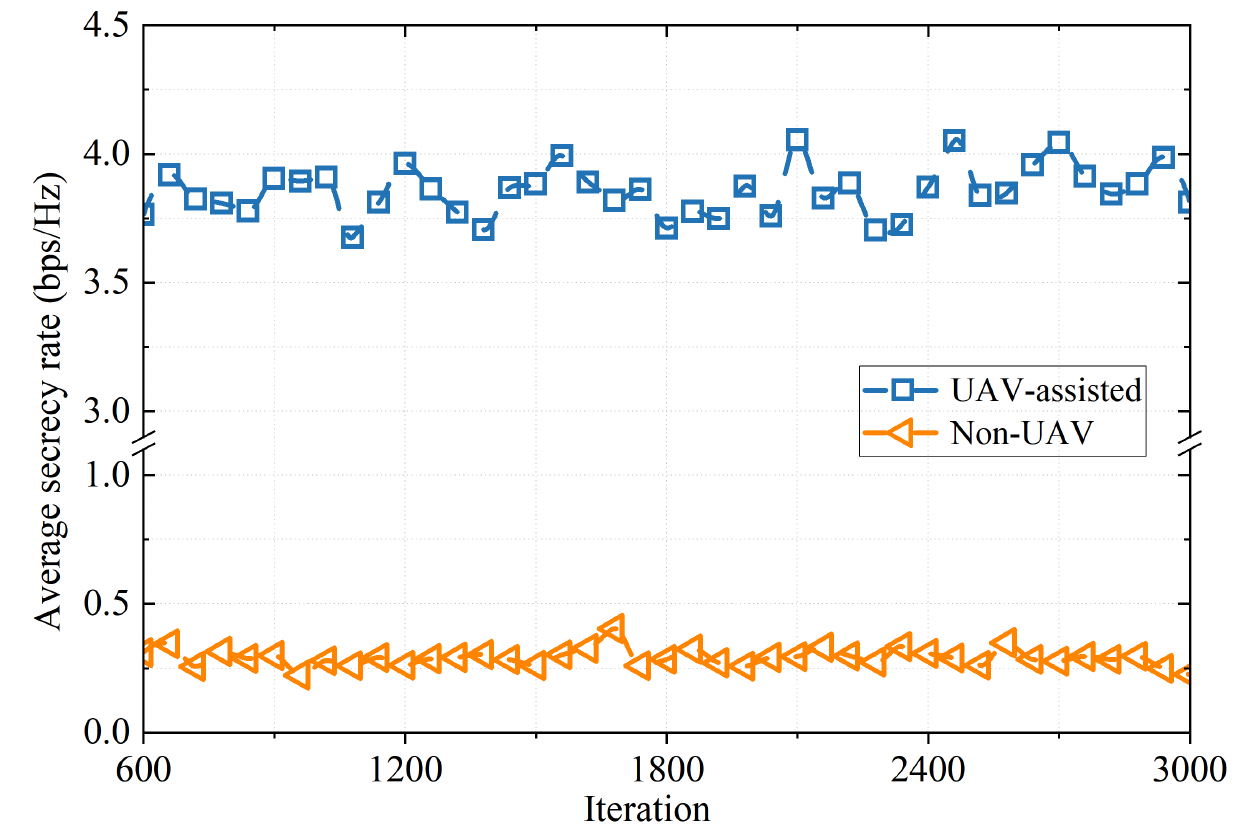}
\caption{Average secrecy rates obtained by the UAV-assisted and non-UAV approaches.}
\label{fig:Non-UAV}
\end{figure}

{\color{black}
\subsubsection{Comparisons with Theoretical Optimal Secrecy Rate}

\par We compare the security performance of LEO satellite-maritime communications between our approach and a theoretical scenario. Fig.~\ref{optimal_secracy_rate} illustrates the secrecy rate obtained by our approach in comparison to the theoretical optimal secrecy rate during an episode. It can be seen that our method approaches the theoretical optimal secrecy rate. Note that our approach solves a multi-objective optimization problem, which requires trade-offs between secrecy rate and energy consumption of the UAV. Thus, our approach achieves a near-optimal secrecy rate, offering a practical and valuable solution for the considered scenario.}

\subsubsection{Comparisons with Non-UAV LEO Satellite-Maritime Communications}

\par We compare the security performance of LEO satellite-maritime communications by using UAV-assisted and non-UAV approaches. Fig. \ref{fig:Non-UAV} presents the average secrecy rates obtained by both approaches. As can be seen, the UAV-assisted approach can consistently maintain superior secrecy rates, ensuring reliable communications. In contrast, the non-UAV approach struggles to reach a comparable secrecy rate. This demonstrates the superior performance of the UAV-assisted approach in achieving secure LEO satellite-maritime communications.

\subsubsection{Comparison with Other Algorithms} 

\par
\begin{figure}
\centering
\includegraphics[width=3.5in]{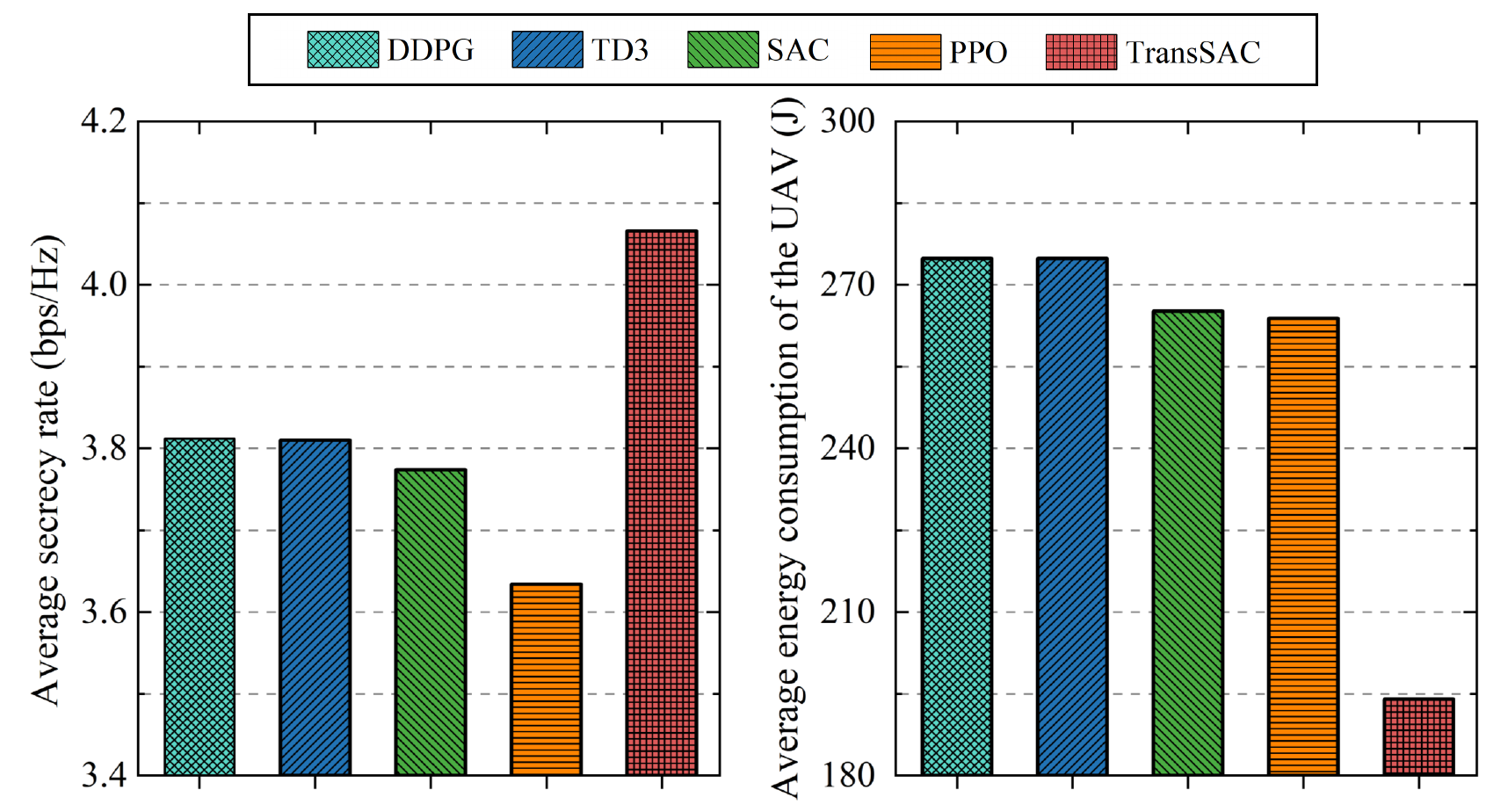}
\caption{The optimization objective values obtained by different algorithms.}
\label{fig:number_value}
\end{figure}

 Fig. \ref{fig:number_value} gives the optimization objective values obtained by different algorithms. Specifically, the TransSAC algorithm has an optimal average secrecy rate, achieving secure communications. Moreover, the average energy consumption of the UAV of the TransSAC algorithm is considerably lower, which is crucial in maritime environments where recharging UAVs is challenging. Therefore, the reliable security and minimal energy consumption make TransSAC a practical and effective algorithm. Notably, since TD3 is an improvement on the DDPG, the average energy consumption of the UAV of TD3 has a minimal enhancement in DDPG, resulting in similar values for both.

\par
\begin{figure}
\centering
\includegraphics[width=3.5in]{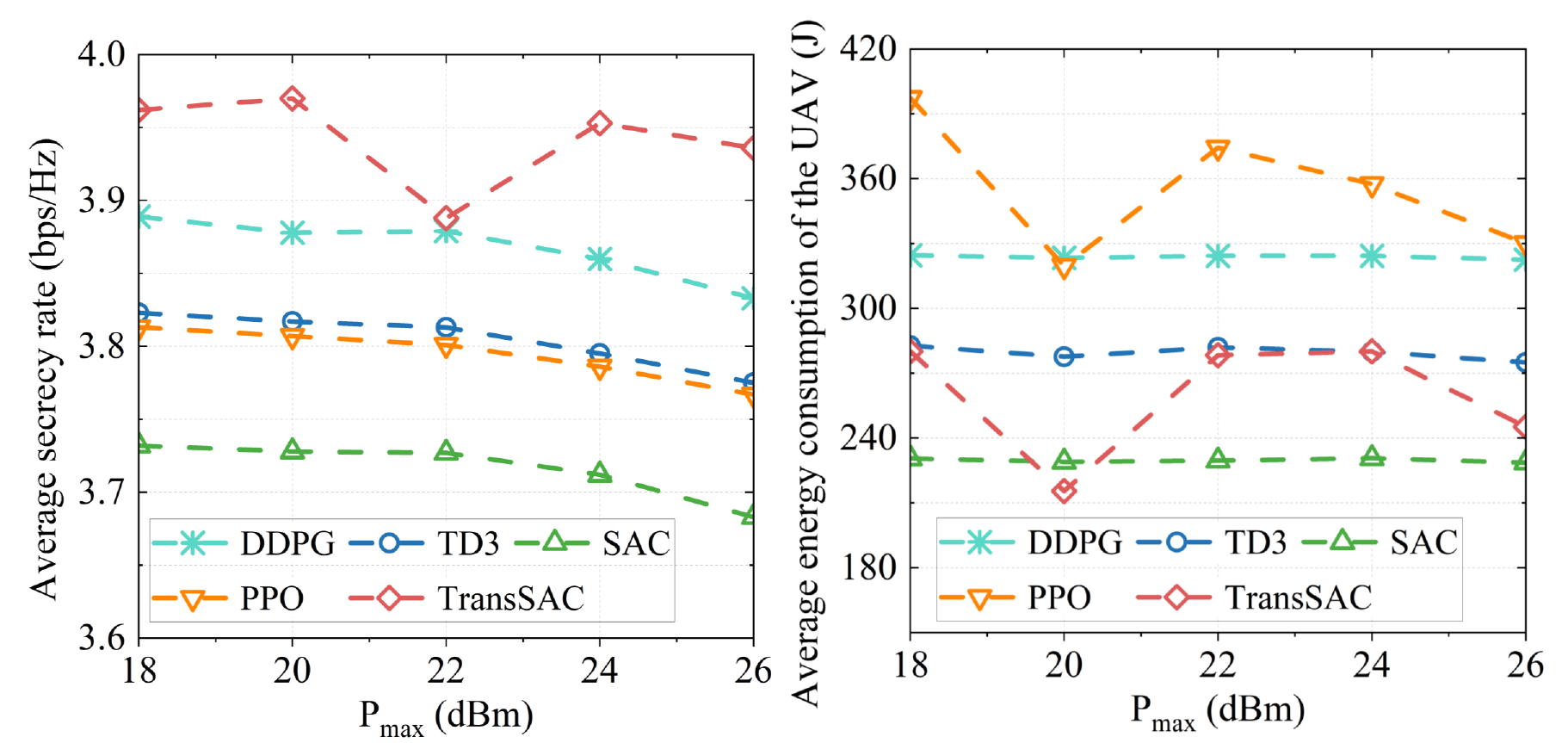}
\caption{Impact of constraint changes on optimization objectives obtained by all algorithms with various $P_{max}$.}
\label{fig:compareP}
\end{figure}

\begin{figure}
\centering
\includegraphics[width=3.5in]{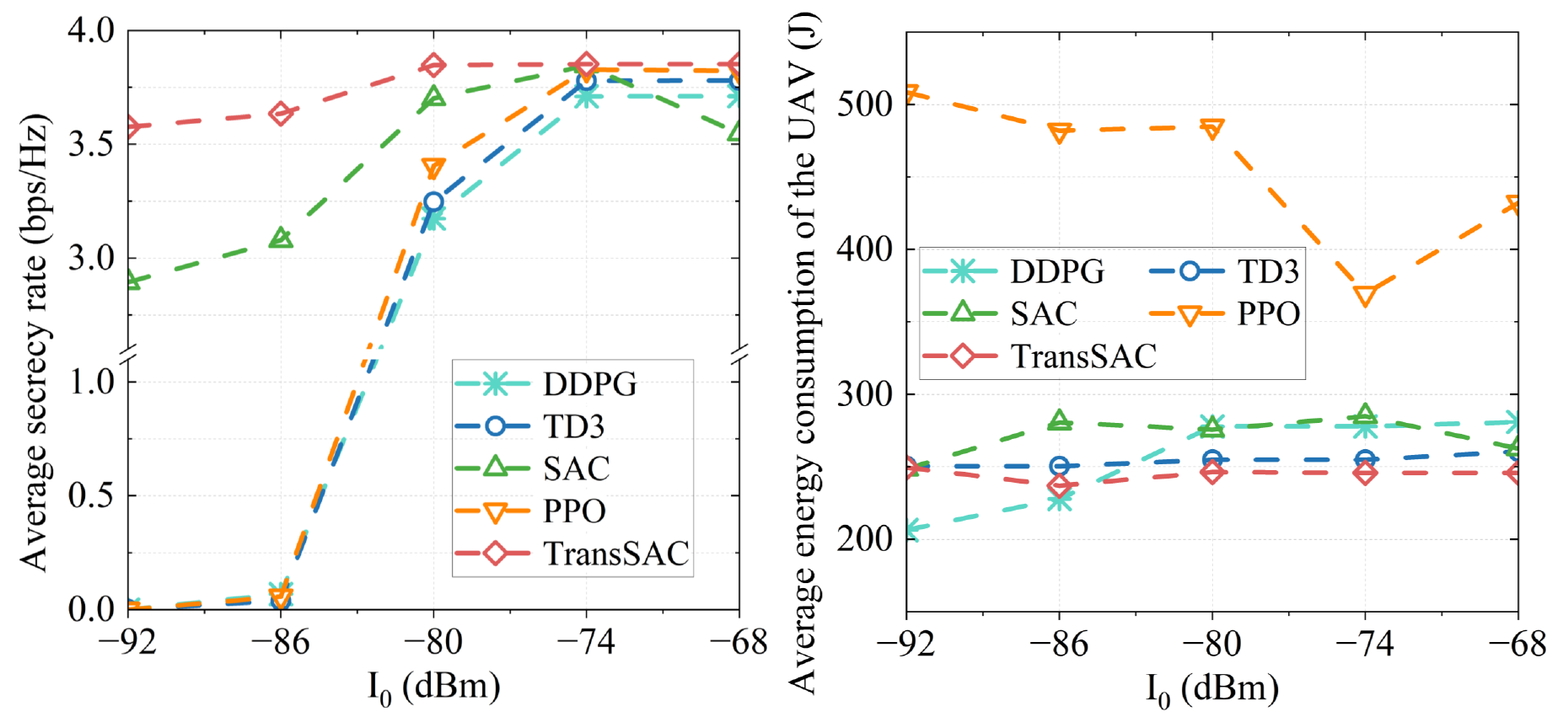}
\caption{Impact of constraint changes on optimization objectives obtained by all algorithms with various $I_{0}$.}
\label{fig:compareI0}
\end{figure}

 Furthermore, we consider determining suitable constraints in the MDP. Fig. \ref{fig:compareP} compares the optimization objective values obtained by different algorithms at various $P_{max}$ values. As can be seen, the average secrecy rates gradually decrease as $P_{max}$ increases. When $P_{max}$ reaches 22, the secrecy rate of TransSAC drops sharply. Moreover, the average energy consumption values of PPO and TransSAC decrease significantly when $P_{max}$ is 20, while the other algorithms show little sensitivity to changes in $P_{max}$. The observations suggest that $P_{max}$ value of 20 is a suitable maximum transmit power to balance objective performance. In addition, Fig. \ref{fig:compareI0} illustrates the optimization objectives of all algorithms at different $I_{0}$ values. Clearly, the average secrecy rates and energy consumption of the UAV are optimal for most algorithms when $I_{0}$ is -74, which serves as an appropriate maximum interference power, and the TransSAC algorithm demonstrates superior performance. Notably, when $I_{0}$ is below -86, the secrecy rates of some algorithms approach 0, indicating ineffective UAV jamming. Therefore, we identify suitable constraints for the MDP. Moreover, the TransSAC algorithm can maximize the secrecy rate and minimize the energy consumption of the UAV under the constraints, further confirming its effectiveness.

\begin{figure}
\centering
\includegraphics[width=3in]{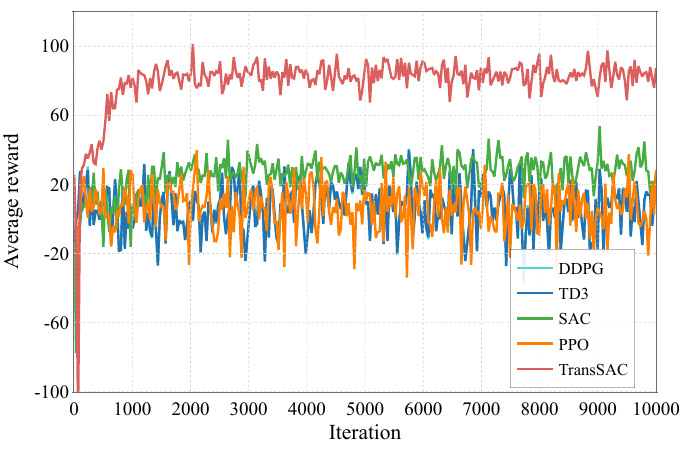}
\caption{Convergence performance obtained by different algorithms.}
\label{fig:reward}
\end{figure}

\subsubsection{Convergence Performance}

\par The convergence performance is a key metric for evaluating DRL, reflecting the ability to stabilize and reach optimal solutions over time. Fig. \ref{fig:reward} indicates the convergence performance of different algorithms. Clearly, the converged TransSAC obtains significantly higher average rewards than those of the other algorithms, demonstrating the ability to learn effective strategies. Notably, TransSAC converges relatively slowly (around 1000 iterations). This is because the self-attention mechanism in the transformer captures more complex features, making the training process more difficult. However, the longer convergence time is acceptable, as the significant improvements in the optimization objectives demonstrate that this trade-off is worthwhile.

%
{\color{black}
\section{Discussion} 
\label{sec:Discussion}

\par In this section, we further discuss the performance of the proposed approach, and the details are as follows:
\begin{itemize}
    \item \textit{The Extended Scenarios of Our Approach:} Our approach can be extended to multiple UAV cases, achieving enhanced security performance and expanding maritime security coverage capabilities, and the details are presented in Appendix A.

    \item \textit{Indicative Simulation Scenario:} We present an indicative simulation scenario, including the trajectories of the LEO satellite, vessels, and the UAV, as well as the performance of key parameters, to enhance the clarity of our results. The detailed analysis and simulations are shown in Appendix B.

    \item \textit{Fulfillment of Constraints of the SSMCMOP:} We elaborate that our approach and various algorithms all satisfy the constraints of the SSMCMOP, and the detailed discussion is provided in Appendix C.

    \item \textit{Implementation of Real-time Data Transmission of Our Approach:} We demonstrate that our DRL-based framework with the offline training and online execution approach can optimize real-time transmission rates by adapting to changing environmental states. Moreover, we use a Raspberry Pi to confirm that our approach can achieve real-time data transmission. See Appendix D for details.
    
\end{itemize}
}

%
\section{Conclusion} 
\label{sec:conclusion}

\par This work has studied secure LEO satellite-maritime communications by low-altitude UAV friendly-jamming. We have considered the system that utilizes a UAV as a low-altitude platform to send jamming signals, mitigating the risk of satellite-transmitted data being eavesdropped by the illegitimate vessel. Considering the conflicting objectives, we have formulated the SSMCMOP to maximize the secrecy rate and minimize the energy consumption of the UAV simultaneously. To tackle the dynamic and NP-hard problem, we have reformulated it into an MDP. {\color{black}Then, we have proposed the TransSAC algorithm, a GenAI-enabled DRL approach that integrates transformer and MAB strategies to capture long-term dependencies and diversely explore weights. Simulation results have shown that the TransSAC algorithm outperforms other comparative methods and algorithms, achieving maximum secrecy rate with minimal energy consumption of the UAV.} Additionally, we have determined the suitable constraint values for the MDP. {\color{black}This work can be extended by incorporating the imperfect positions of eavesdropping vessels and leveraging a DRL-based framework to predict the position of the illegitimate user, which will be explored in future work.}

\bibliographystyle{IEEEtran}
\bibliography{myref}{}

\vspace{-18mm}
\begin{IEEEbiography}
[{\includegraphics[width=1in,height=1.25in,clip,keepaspectratio]{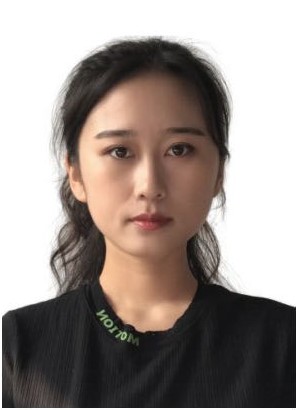}}]
{Jiawei Huang} received a BS degree in Software Engineering from Dalian Jiaotong University, and a MS degree in Software Engineering from Jilin University in 2019 and 2024, respectively. She is currently studying Computer Science at Jilin University to get a Ph.D. degree. Her current research interests are UAV networks and optimization.
\end{IEEEbiography}

\vspace{-18mm}
\begin{IEEEbiography}
[{\includegraphics[width=1in,height=1.25in,clip,keepaspectratio]{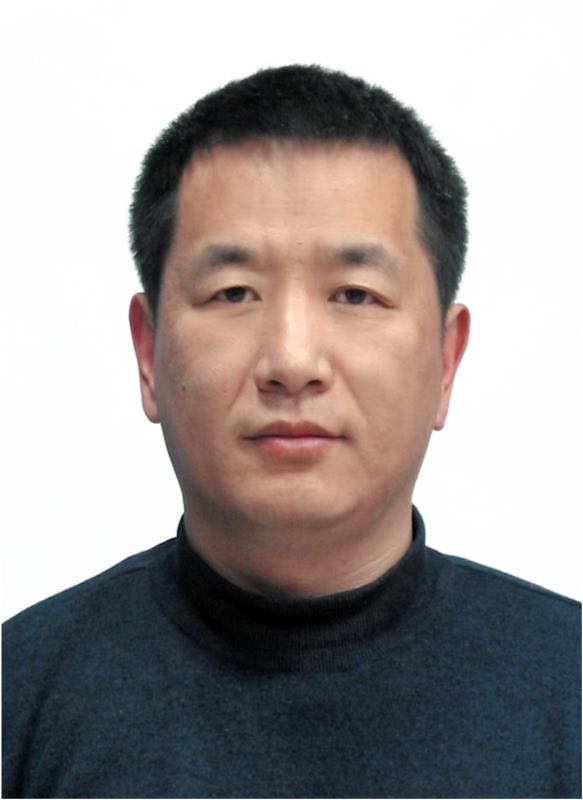}}]
{Aimin Wang} received Ph.D. degree in Communication and Information System from Jilin University, Changchun, China, in 2004. He is currently a professor at Jilin University. His research interests are wireless sensor networks and QoS for multimedia transmission.
\end{IEEEbiography}

\vspace{-18mm}
\begin{IEEEbiography}
[{\includegraphics[width=1in,height=1.25in,clip,keepaspectratio]{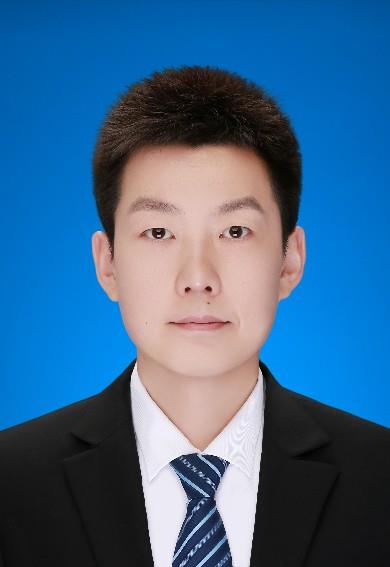}}]
{Geng Sun} (Senior Member, IEEE) received the B.S. degree in communication engineering from Dalian Polytechnic University, and the Ph.D. degree in computer science and technology from Jilin University, in 2011 and 2018, respectively. He was a Visiting Researcher with the School of Electrical and Computer Engineering, Georgia Institute of Technology, USA. He is a Professor in the College of Computer Science and Technology at Jilin University. Currently, he is working as a visiting scholar at the College of Computing and Data Science, Nanyang Technological University, Singapore. He has published over 100 high-quality papers, including IEEE TMC, IEEE JSAC, IEEE/ACM ToN, IEEE TWC, IEEE TCOM, IEEE TAP, IEEE IoT-J, IEEE TIM, IEEE INFOCOM, IEEE GLOBECOM, and IEEE ICC. He serves as the Associate Editors of IEEE Communications Surveys \& Tutorials, IEEE Transactions on Communications, IEEE Transactions on Vehicular Technology, IEEE Transactions on Network Science and Engineering, IEEE Transactions on Network and Service Management and IEEE Networking Letters. He serves as the Lead Guest Editor of Special Issues for IEEE Transactions on Network Science and Engineering, IEEE Internet of Things Journal, IEEE Networking Letters. He also serves as the Guest Editor of Special Issues for IEEE Transactions on Services Computing, IEEE Communications Magazine, and IEEE Open Journal of the Communications Society. His research interests include Low-altitude Wireless Networks, UAV communications and Networking, Mobile Edge Computing (MEC), Intelligent Reflecting Surface (IRS), Generative AI and Agentic AI, and deep reinforcement learning.

\end{IEEEbiography}

\begin{IEEEbiography}
[{\includegraphics[width=1in,height=1.25in,clip,keepaspectratio]{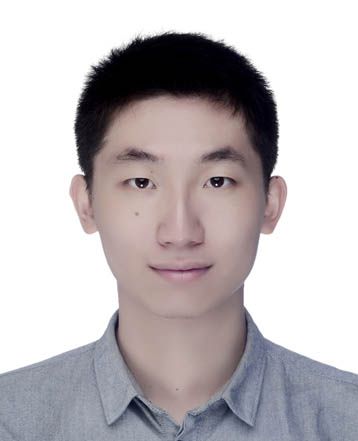}}]
{Jiahui Li} (Member, IEEE) received his B.S. in Software Engineering, and M.S. and Ph.D. in Computer Science and Technology from Jilin University, Changchun, China, in 2018, 2021, and 2024, respectively. He was a visiting Ph.D. student at the Singapore University of Technology and Design (SUTD). He currently serves as an assistant researcher in the College of Computer Science and Technology at Jilin University. His current research focuses on integrated air-ground networks, UAV networks, wireless energy transfer, and optimization.

\end{IEEEbiography}
\vspace{-18mm}
\begin{IEEEbiography}
[{\includegraphics[width=1in,height=1.25in,clip,keepaspectratio]{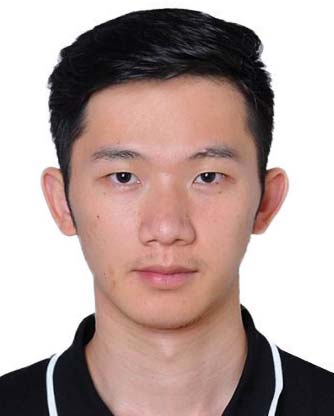}}]{Jiacheng Wang} received the Ph.D. degree from the School of Communication and Information Engineering, Chongqing University of Posts and Telecommunications, Chongqing, China. He is currently a Research Associate in computer science and engineering with Nanyang Technological University, Singapore. His research interests include wireless sensing, semantic communications, and metaverse.
\end{IEEEbiography}
\vspace{-18mm}
\begin{IEEEbiography}
[{\includegraphics[width=1in,height=1.25in,clip,keepaspectratio]{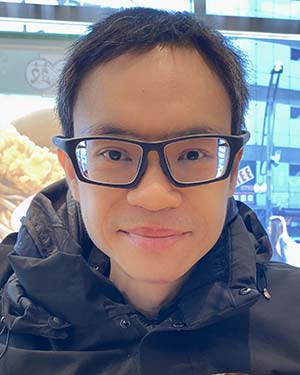}}]{Dusit Niyato} (M'09-SM'15-F'17) is a professor in the College of Computing and Data Science, at Nanyang Technological University, Singapore. He received B.Eng. from King Mongkuts Institute of Technology Ladkrabang (KMITL), Thailand and Ph.D. in Electrical and Computer Engineering from the University of Manitoba, Canada. His research interests are in the areas of mobile generative AI, edge general intelligence, quantum computing and networking, and incentive mechanism design.
\end{IEEEbiography}

\vspace{-18mm}
\begin{IEEEbiography}
[{\includegraphics[width=1in,height=1.25in,clip,keepaspectratio]{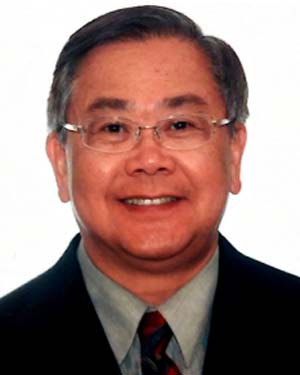}}] {Victor C. M. Leung} (Life Fellow, IEEE) is currently a distinguished professor of computer science and software engineering with Shenzhen University, China. He is also an emeritus professor of electrial and computer engineering and the Director with the Laboratory for Wireless Networks and Mobile Systems, University of British Columbia. He has coauthored more than 1300 journal/conference papers and book chapters, and has been named in the current Clarivate Analytics list of Highly Cited Researchers. His research interests include the broad areas of wireless networks and mobile systems. Dr. Leung is also on the editorial boards of IEEE Transactions on Green Communications and Networking, IEEE Transactions on Cloud Computing, IEEE Access, and several other journals. He was the recipient of the IEEE Vancouver Section Centennial Award, 2011 UBC Killam Research Prize, 2017 Canadian Award for Telecommunications Research, 2018 IEEE TCGCC Distinguished Technical Achievement Recognition Award, and has coauthored papers that were the recipient of the 2017 IEEE ComSoc Fred W. Ellersick Prize, 2017 IEEE Systems Journal Best Paper Award, 2018 IEEE CSIM Best Journal Paper Award, and 2019 IEEE TCGCC Best Journal Paper Award. He is also the Life Fellow of IEEE, and a Fellow of the Royal Society of Canada, Canadian Academy of Engineering, and Engineering Institute of Canada.
\end{IEEEbiography}

\end{document}